\documentclass[useAMS]{mn2e}
\usepackage{times}
\usepackage{epsfig}
\usepackage{amsmath}

\title[$\epsilon$~Eri]{Modelling the chromosphere and transition region
of $\epsilon$~Eri (K2 V)}
\author[S. A. Sim, C. Jordan]{S. A. Sim$^{1,2}$\thanks{s.sim@imperial.ac.uk}, C. Jordan$^{2}$ \\
$^1$Astrophysics Group, Imperial College London,
Blackett Laboratory, Prince Consort Road, London, SW7 2AZ, UK\\
$^2$Department of Physics (The Rudolf Peierls Centre for Theoretical Physics), University of Oxford,
1 Keble Road, OX1 3NP, UK}

\date{\today}

\pagerange{\pageref{firstpage}--\pageref{lastpage}}
\pubyear{2005}

\begin{document}
\maketitle
\label{firstpage}

\begin{abstract}
Measurements of ultraviolet line fluxes from Space Telescope Imaging
Spectrograph and {\it Far-Ultraviolet Spectroscopic Explorer} 
spectra of the K2-dwarf
$\epsilon$~Eri are reported. 
These are used to develop new emission measure distributions
and semi-empirical atmospheric models for the chromosphere and
lower transition region of the star. These models are the most detailed
constructed to date for a main-sequence star other than the Sun.
New ionisation balance calculations,
which account for the effect of finite density on dielectronic
recombination rates, are presented for carbon, nitrogen, oxygen
and silicon. The results of these calculations are significantly
different from the standard Arnaud \& Rothenflug ion balance, 
particularly for alkali-like ions.
The new atmospheric models are used to place constraints on possible
First Ionisation Potential (FIP) related abundance variations in the
lower atmosphere and to discuss limitations of
single-component models for the interpretation of certain
optically thick line fluxes.
\end{abstract}


\section{Introduction}

Although it has long been known that cool main-sequence stars have hot
outer atmospheres the 
question of how these regions are heated remains
unanswered.
This is, in part, due to the
complexity of stellar atmospheres arising from the
interplay of sub-photospheric convection, rotation and 
magnetic fields --  
in order to distinguish
between different physical processes quantitative analyses 
and models are
required.

Atmospheric models fall into two broad categories, those
which are constructed from theory 
(e.g. those developed by Fontenla, 
Avrett \& Loeser 1993, 2002 or Cuntz et al. 1999)
and those which
are constructed semi-empirically, based on a combination of
observations and a few elementary physical assumptions
(e.g. the Vernazza, Avrett \& Loeser 1973, 1976, 1981 solar models). 
Semi-empirical models
are vital since it is only by
comparison with properties deduced from observations that theoretical models
can be tested.

In this paper new semi-empirical models are developed for the 
chromosphere and lower transition region of the K2-dwarf
$\epsilon$~Eri. This 
complements our recent studies of the outer atmosphere in which
the mean electron pressure was measured (Jordan et al. 2001a);
line widths were measured and used to investigate the non-thermal 
velocity fields within the atmosphere and place constraints upon the
possibility of coronal heating by wave dissipation (Sim \& Jordan 
2003a); and emission measure distributions were used to derive new 
information
about the area occupied by emitting material in the
upper transition region (Sim \& Jordan 2003b).

Previously, semi-empirical models have been constructed for $\epsilon$~Eri by
Kelch (1978), Simon, Kelch \& Linsky (1980) and 
Thatcher, Robinson \& Rees (1991).
The Kelch (1978) model was based on observations of the Ca~{\sc ii} and
Mg~{\sc ii} resonance lines and described the upper photosphere and
chromosphere (extending to an electron temperature 
$\log T_{e}[\mbox{K}] =4.3$). Simon et al. (1980) used ultraviolet observations
of Si~{\sc ii}, Si~{\sc iii} and C~{\sc ii} recorded with the {\it 
International Ultraviolet Explorer} ({\it IUE}) 
to develop the upper parts of
the Kelch (1978) model. Their model extends to $\log T_{e}[\mbox{K}] =4.7$.
Thatcher et al. (1991) used infrared lines of Ca~{\sc ii}, the Na D
doublet and lines from the H Balmer series to further develop the chromospheric
part of the Kelch (1978) model and extended the model to 
$\log T_{e}[\mbox{K}] =5.6$.

The new models presented here are based on ultraviolet observations
made with the {Space Telescope Imaging Spectrograph} (STIS) on board the
{\it Hubble Space Telescope} and with the {\it Far-Ultraviolet Spectroscopic
Explorer} ({\it FUSE}) satellite (see Section 2). The high signal-to-noise
ratio, high spectral resolution and large combined spectral coverage of 
these data are ideal for quantitative modelling;
they provide much tighter constraints on the model 
than did the {\it IUE} data and allow much more robust conclusions to be drawn.
In particular, the improved measurements of the transition region pressure,
together with radiative transfer calculations,
lead to more reliable models from the photosphere to the transition region.

An important limitation of the available spectroscopic data is that, in 
common with all such observations of main-sequence stars other than the Sun,
they contain no direct information on the spatial inhomogeneities
(because the source is unresolved). Although we have been able to deduce 
information about the inhomogeneity in the upper transition
region (Sim \& Jordan 2003b), the same method cannot yet be applied to the
deeper parts of the atmosphere since the local means of energy deposition is 
not understood. Here, for the most part,
the simplest assumption, that of homogeneity, is made -- thus
the models are representative of the ``average''
atmosphere. 
Although the interpretation of ``average'' models requires
considerable care, they are extremely important since they
contain the most information that can be extracted directly from the data
alone and theoretical models must be consistent with the mean properties.
Two-component theoretical models of a K2~V star chromosphere have been made by 
Cuntz et al. (1999), using empirical relations between the area covered by 
magnetic fields and stellar rotation periods. 
We also derive a two-component by applying the emitting area
that we found at $10^5$~K (Sim \& Jordan 2003b) as an upper
limit to regions at lower temperatures -- solar observations show that the
area occupied by supergranulation cells remains constant between 
$\sim 2 \times 10^4$~K and $10^5$~K, before decreasing in the chromosphere.

Another limitation to the modelling 
is the accuracy of the atomic data used -- 
for many of the measured line fluxes, the observational
uncertainty is small compared to possible errors in the atomic rates.
In particular, the standard sets of ionisation balance calculations used for
transition region lines are inadequate for modelling to the precision of the
{STIS} data. In this paper a new ionisation balance calculation is presented
in which rates that account for variations in the dielectronic
recombination rate with density are used; although not definitive, this
calculation highlights the need for further work on recombination rates
at non-zero density.

The ultraviolet observations and measurements of the emission line
fluxes are discussed in Section 2. In Section 3 the new ionisation balance
calculations are discussed. They are used in the construction of
emission measure distributions (EMDs) in Section 4. A
preliminary single-component 
atmospheric model is developed in Section 5 and improvements
to this model are discussed in Section 6. 
In Section~7, the simple two-component model is presented and 
conclusions are drawn in Section 8.

\section{Observations}

\subsection{Flux measurements}

The models developed are based on measurements of
fluxes for a variety of emission lines in {STIS} and {\it FUSE} spectra of
$\epsilon$~Eri.
Table~1 gives the measured fluxes for lines in the {STIS} 
spectra (see Jordan et al. 2001b for details of our STIS spectra). These
were obtained by direct integration across line profiles and, where possible,
by performing single-Gaussian fits. (The line widths and shifts deduced from 
Gaussian fits have been published elsewhere: Sim \& Jordan 2003a.) The
errors given are tolerances based on the signal-to-noise, the uncertainty
in the background level and, where applicable, the goodness of the Gaussian
fits. Four of the lines listed in Table~1 are affected by blending. Three
of these (O~{\sc iv} 1404.8~\AA, Si~{\sc ii} 2335.3~\AA~and S~{\sc iv} 
1404.8~\AA)
have been discussed by Jordan et al. (2001a). The other, 
Si~{\sc ii} 1260.4~\AA, is
suspected to be a close blend with C~{\sc i} 1260.6~\AA. 
The Al~{\sc iii} $^2$S$_{1/2}$ -- $^2$P$_{1/2}$ line at 1862.8~\AA~ it too
weak to measure. This suggests that the Al~{\sc iii} 1854.7-\AA~ line is
also blended, probably with an iron line. 
(In the optically thin limit the
Al~{\sc iii} 1854.7-\AA~line should have double the flux of the
1862.8-\AA~line, but the observed 1854.7-\AA~feature is more than twice as
strong as that at 1862.8~\AA.)

\begin{table*}
\caption{Emission line fluxes measured from the $\epsilon$~Eri {\it STIS} spectra. 
$\lambda_{0}$ is the rest wavelength. ``bl.'' indicates that the line
is part of a blend: in such cases the {\it total} measured flux in the blend is
given as an upper limit to the line flux.
The rest wavelengths are taken from Griesmann \& Kling (2000) [C~{\sc iv} and Si~{\sc iv}],
Kaufman \& Martin (1993) [S~{\sc iv}] and Kurucz \& Bell (1995) [all other lines].
}
\begin{tabular}{lccc}\hline
Ion & Transition & $\lambda_{0}$ & Flux at Earth \\ 
&&(\AA)&($10^{-13}$ ergs cm$^{-2}$ s$^{-1}$)\\\hline
H {\sc i} & 1s $^{2}$S -- 2p $^{2}$P & 1215.671 & $160 \pm 1$ \\
C {\sc ii} & 2s$^{2}$2p $^{2}$P$_{1/2}$ -- 2s2p$^{2}$ $^{2}$D$_{3/2}$ &
1334.532 & $1.88\pm0.06$ \\
C {\sc ii} & 2s$^{2}$2p $^{2}$P$_{3/2}$ -- 2s2p$^{2}$ $^{2}$D$_{5/2,3/2}$ &
1335.68 bl. & $4.84\pm0.10$ \\
C {\sc iii} & 2s2p $^3$P$_1$ -- 2p$^2$ $^3$P$_2$  &
1174.933 & $0.41\pm0.06$ \\
C {\sc iii} & 2s2p $^3$P$_0$ -- 2p$^2$ $^3$P$_1$  &
1175.263 & $0.41\pm0.06$ \\
C {\sc iii} & 2s2p $^3$P$_{2,1}$ -- 2p$^2$ $^3$P$_{2,1}$  &
1175.69 bl. & $1.49\pm0.06$ \\
C {\sc iii} & 2s2p $^3$P$_1$ -- 2p$^2$ $^3$P$_0$  &
1175.987 & $0.41\pm0.06$ \\
C {\sc iii} & 2s2p $^3$P$_2$ -- 2p$^2$ $^3$P$_1$  &
1176.370 & $0.46\pm0.06$ \\
C {\sc iv}  & 2s $^{2}$S$_{1/2}$ -- 2p $^{2}$P$_{3/2}$   &
1548.204 & $5.59\pm0.11$ \\
C {\sc iv}  & 2s $^{2}$S$_{1/2}$ -- 2p $^{2}$P$_{1/2}$   &
1550.781 & $2.82\pm0.07$ \\
N {\sc iv}  & 2s$^{2}$ $^{1}$S$_{0}$ -- 2s2p $^{3}$P$_{1}$   &
1486.496 & $0.045\pm0.015$ \\
N {\sc v}  & 2s $^{2}$S$_{1/2}$ -- 2p $^{2}$P$_{3/2}$   &
1238.821& $0.94\pm 0.03$ \\
N {\sc v}  & 2s $^{2}$S$_{1/2}$ -- 2p $^{2}$P$_{1/2}$   &
1242.804& $0.46\pm0.02$ \\
O {\sc iii} & 2s$^{2}$2p$^{2}$ $^{3}$P$_{1}$ -- 2s2p$^{3}$ $^{5}$S$_{2}$ & 
1660.809 & $0.058\pm0.015$ \\
O {\sc iii} & 2s$^{2}$2p$^{2}$ $^{3}$P$_{2}$ -- 2s2p$^{3}$ $^{5}$S$_{2}$ & 
1666.150 & $0.121\pm0.015$ \\
O {\sc iv} & 2s$^{2}$2p $^{2}$P$_{1/2}$ -- 2s2p$^{2}$ $^{4}$P$_{1/2}$ &
1399.780& $0.054\pm0.005$ \\
O {\sc iv} & 2s$^{2}$2p $^{2}$P$_{3/2}$ -- 2s2p$^{2}$ $^{4}$P$_{5/2}$ &
1401.157 & $0.181\pm0.005$ \\
O {\sc iv} & 2s$^{2}$2p $^{2}$P$_{3/2}$ -- 2s2p$^{2}$ $^{4}$P$_{3/2}$ &
1404.806 & $0.069\pm0.005$ bl. \\
O {\sc iv} & 2s$^{2}$2p $^{2}$P$_{3/2}$ -- 2s2p$^{2}$ $^{4}$P$_{1/2}$ &
1407.382 & $0.052\pm0.005$ \\
O {\sc v} & 2s$^2$ $^{1}$S$_{0}$ -- 2s2p $^{3}$P$_{1}$ &
1218.344 & $0.68\pm0.04$ \\
O {\sc v} & 2s2p $^{1}$P$_{1}$ -- 2p$^{2}$ $^{1}$D$_{2}$ &
1371.296 & $0.105\pm0.005$ \\
Mg {\sc ii} & 3s $^{2}$S$_{1/2}$ -- 3p $^{2}$P$_{3/2}$&
2796.352 & $305\pm6$\\
Mg {\sc ii} & 3s $^{2}$S$_{1/2}$ -- 3p $^{2}$P$_{1/2}$&
2803.531 & $232\pm5$\\
Al {\sc iii} & 3s $^{2}$S$_{1/2}$ -- 3p $^2$P$_{3/2}$ &
1854.716 & $0.51\pm0.06$ bl? \\
Si {\sc ii} & 3s$^{2}$3p $^{2}$P$_{1/2}$ -- 3s3p$^{2}$ $^{2}$P$_{3/2}$ &
1190.416 & $0.10\pm0.03$ \\
Si {\sc ii} & 3s$^{2}$3p $^{2}$P$_{1/2}$ -- 3s3p$^{2}$ $^{2}$P$_{1/2}$ &
1193.290 & $0.09\pm0.04$ \\
Si {\sc ii} & 3s$^{2}$3p $^{2}$P$_{3/2}$ -- 3s3p$^{2}$ $^{2}$P$_{3/2}$ &
1194.500 & $0.18\pm0.03$ \\
Si {\sc ii} & 3s$^{2}$3p $^{2}$P$_{3/2}$ -- 3s3p$^{2}$ $^{2}$P$_{1/2}$ &
1197.394 & $0.10\pm0.04$ \\
Si {\sc ii} &  3p $^{2}$P$_{1/2}$ -- 3d $^{2}$D$_{3/2}$ &
1260.422 & $0.191\pm0.03$ bl. \\
Si {\sc ii} & 3p $^{2}$P$_{3/2}$ -- 3d $^{2}$D$_{5/2}$ &
1264.738 & $0.505\pm0.05$ \\
Si {\sc ii} & 3p $^{2}$P$_{3/2}$ -- 3d $^{2}$D$_{3/2}$ &
1265.002 & $0.172\pm0.03$ \\
Si {\sc ii} & 3s$^{2}$3p $^{2}$P$_{1/2}$ -- 3s3p$^{2}$ $^{2}$S$_{1/2}$ &
1304.370 & $0.133\pm0.008$\\
Si {\sc ii} & 3s$^{2}$3p $^{2}$P$_{3/2}$ -- 3s3p$^{2}$ $^{2}$S$_{1/2}$ &
1309.276 & $0.222\pm0.010$\\
Si {\sc ii} & 3p $^{2}$P$_{1/2}$ -- 4s $^{2}$S$_{1/2}$ &
1526.707 & $0.508\pm0.005$ \\
Si {\sc ii} & 3p $^{2}$P$_{3/2}$ -- 4s $^{2}$S$_{1/2}$ &
1533.431 & $0.607\pm0.005$ \\
Si {\sc ii} & 3s$^{2}$3p $^{2}$P$_{1/2}$ -- 3s3p$^{2}$ $^{2}$D$_{3/2}$ &
1808.013 & $6.39\pm0.06$ \\
Si {\sc ii} & 3s$^{2}$3p $^{2}$P$_{3/2}$ -- 3s3p$^{2}$ $^{2}$D$_{5/2}$ &
1816.928 & $10.42\pm0.07$ \\
Si {\sc ii} & 3s$^{2}$3p $^{2}$P$_{3/2}$ -- 3s3p$^{2}$ $^{2}$D$_{3/2}$ &
1817.451 & $3.20\pm0.06$ \\
Si {\sc ii} & 3s$^{2}$3p $^{2}$P$_{1/2}$ -- 3s3p$^{2}$ $^{4}$P$_{1/2}$ &
2335.123 & $3.31\pm0.07$ \\
Si {\sc ii} & 3s$^{2}$3p $^{2}$P$_{3/2}$ -- 3s3p$^{2}$ $^{4}$P$_{5/2}$ &
2335.321 & $2.23\pm0.07$ bl.\\
Si {\sc ii} & 3s$^{2}$3p $^{2}$P$_{3/2}$ -- 3s3p$^{2}$ $^{4}$P$_{1/2}$ &
2350.891 & $3.04\pm0.07$\\
Si {\sc iii} & 3s$^{2}$ $^{1}$S$_{0}$ -- 3s3p $^{1}$P$_{1}$ &
1206.500 & $3.60\pm0.10$\\
Si {\sc iii} & 3s3p $^{3}$P$_{1}$ -- 3p$^{2}$ $^{3}$P$_{2}$ &
1294.545 & $0.052\pm0.005$ \\
Si {\sc iii} & 3s3p $^{3}$P$_{0}$ -- 3p$^{2}$ $^{3}$P$_{1}$ &
1296.726 & $0.041\pm0.005$ \\
Si {\sc iii} & 3s3p $^{3}$P$_{2,1}$ -- 3p$^{2}$ $^{3}$P$_{2,1}$ &
1298.93 bl. & $0.167\pm0.005$ \\
Si {\sc iii} & 3s3p $^{3}$P$_{1}$ -- 3p$^{2}$ $^{3}$P$_{0}$ &
1301.149 & $0.028\pm0.005$ \\
Si {\sc iii} & 3s3p $^{3}$P$_{2}$ -- 3p$^{2}$ $^{3}$P$_{1}$ &
1303.323 & $0.062\pm0.005$ \\
Si {\sc iii} & 3s$^{2}$ $^{1}$S$_{0}$ -- 3s3p $^{3}$P$_{1}$ &
1892.030 & $1.25\pm0.05$ \\
Si {\sc iv} & 3s $^{2}$S$_{1/2}$ -- 3p $^{2}$P$_{3/2}$   &
1393.760 & $2.30\pm0.06$ \\
Si {\sc iv} & 3s $^{2}$S$_{1/2}$ -- 3p $^{2}$P$_{1/2}$   &
1402.773 & $1.21\pm0.05$ \\
S {\sc ii} & 3s$^{2}$3p$^{3}$ $^{4}$S$_{3/2}$ -- 3s3p$^{4}$ $^{4}$P$_{1/2}$ &
1250.584 & $0.029\pm0.010$ \\
S {\sc ii} & 3s$^{2}$3p$^{3}$ $^{4}$S$_{3/2}$ -- 3s3p$^{4}$ $^{4}$P$_{3/2}$ &
1253.811 & $0.066\pm0.005$ \\
S {\sc ii} & 3s$^{2}$3p$^{3}$ $^{4}$S$_{3/2}$ -- 3s3p$^{4}$ $^{4}$P$_{5/2}$ &
1259.519 & $0.099\pm0.005$  \\
S {\sc iv} & 3s$^{2}$3p $^{2}$P$_{1/2}$ -- 3s3p$^{2}$ $^{4}$P$_{1/2}$ &
1404.808 & $0.069\pm0.005$ bl. \\
S {\sc iv} & 3s$^{2}$3p $^{2}$P$_{3/2}$ -- 3s3p$^{2}$ $^{4}$P$_{5/2}$ &
1406.016 & $0.032\pm0.007$\\
S {\sc iv} & 3s$^{2}$3p $^{2}$P$_{3/2}$ -- 3s3p$^{2}$ $^{4}$P$_{3/2}$ &
1416.887 & $0.042\pm0.007$\\
Cl {\sc i} & 3p$^5$ $^2$P$_{1/2}$ -- 3p$^{4}$($^3$P)4s $^2$P$_{1/2}$ &
1351.656 & $0.198 \pm 0.005$\\
\hline
\end{tabular}
\end{table*}

It is noteworthy that when the {\it IUE} fluxes
reported for $\epsilon$~Eri by Jordan et al. (1987) are compared with
the new fluxes from {STIS} there is no evidence of significant
differences. Since the {\it IUE} and {STIS} data were recorded seventeen 
years apart this argues against significant long term variability
in the atmosphere.

The {STIS} fluxes are supplemented by measurements from {\it FUSE} spectra
of $\epsilon$~Eri. The {\it FUSE} observations are discussed by
Redfield et al. (2002). The lines in the {\it FUSE} spectrum that are used
here, and our measurements of their 
fluxes, are listed in Table~2. As for the STIS
measurements, the fluxes were obtained by a combination of
direct integration and single-Gaussian fits. The 
{\it FUSE} instrument has four channels which record spectra, 1-LiF, 2-LiF,
1-SiC and 2-Sic.
Most of the {\it FUSE} line
fluxes have been measured from spectra recorded with the 1-LiF channel
(this channel is used to guide the {\it FUSE} line-of-sight). Unfortunately,
the S~{\sc vi} 933.4-\AA, C~{\sc iii} 977.0-\AA~ and 
H~{\sc i} 972.5-\AA~lines do not fall 
in the 1-LiF spectra. Therefore, the fluxes for these lines have been measured
from 1-SiC spectra which are susceptible to losing photons due to channel
drift (Sankrit et al. 2001). Measurements of the strong O~{\sc vi} line
fluxes (which occur in both the 1-LiF and 1-SiC spectra) were used to correct
the 1-SiC line fluxes for channel drift but 
the measurement uncertainties are still largest for the 1-SiC measurements.

\begin{table*}
\caption{Fluxes for emission lines in the $\epsilon$~Eri {\it FUSE} spectra.
Unless otherwise indicated, all lines were measured in the 1-LiF spectra.
$\lambda_0$ is the rest wavelength. The rest wavelengths are taken from
Kurucz \& Bell (1995).}
\begin{tabular}{lccc} \hline
Ion & Transition & $\lambda_{0}$ & Flux at Earth \\ 
&&(\AA)&($10^{-13}$ ergs cm$^{-2}$ s$^{-1}$)\\\hline
H~{\sc i}$^a$ & 1s $^{2}$S -- 4p $^{2}$P & 972.537 & $0.63\pm0.08$ \\ 
H~{\sc i} & 1s $^{2}$S -- 3p $^{2}$P & 1025.722 & $3.66\pm0.12$  \\ 
C~{\sc ii} & 2s2p$^{2}$ $^{4}$P$_{1/2}$ -- 2p$^{3}$ $^{4}$S$_{3/2}$ &
1009.858 & $0.012\pm0.007$\\
C~{\sc ii} & 2s2p$^{2}$ $^{4}$P$_{3/2}$ -- 2p$^{3}$ $^{4}$S$_{3/2}$ &
1010.083 & $0.027\pm0.007$ \\
C~{\sc ii} & 2s2p$^{2}$ $^{4}$P$_{5/2}$ -- 2p$^{3}$ $^{4}$S$_{3/2}$ &
1010.371 & $0.046\pm0.007$ \\
C~{\sc ii} & 2s$^{2}$2p $^{2}$P$_{1/2}$ -- 2s2p$^{2}$ $^{2}$S$_{1/2}$ &
1036.337& $0.120\pm0.014$ \\
C~{\sc ii} & 2s$^{2}$2p $^{2}$P$_{3/2}$ -- 2s2p$^{2}$ $^{2}$S$_{1/2}$ &
1037.018& $0.291\pm0.008$ \\
C~{\sc iii}$^a$ & 2s$^{2}$ $^{1}$S$_{0}$ -- 2s2p $^{1}$P$_{1}$ &
977.020& $4.67\pm0.10$ \\
C {\sc iii} & 2s2p $^3$P$_1$ -- 2p$^2$ $^3$P$_2$  &
1174.933 & $0.435\pm0.015$\\
C {\sc iii} & 2s2p $^3$P$_0$ -- 2p$^2$ $^3$P$_1$  &
1175.263 & $0.425\pm0.015$\\
C {\sc iii} & 2s2p $^3$P$_{2,1}$ -- 2p$^2$ $^3$P$_{2,1}$  &
1175.63 bl. & $1.519\pm0.015$\\
C {\sc iii} & 2s2p $^3$P$_1$ -- 2p$^2$ $^3$P$_0$  &
1175.987 & $0.405\pm0.015$\\
C {\sc iii} & 2s2p $^3$P$_2$ -- 2p$^2$ $^3$P$_1$  &
1176.370 & $0.467\pm0.015$\\
N {\sc iii} & 2s$^{2}$2p $^{2}$P$_{3/2}$ -- 2s2p$^{2}$ $^{2}$D$_{3/2,5/2}$ &
991.54 bl.& $0.33\pm0.08$\\
O {\sc vi} & 2s $^{2}$S$_{1/2}$ -- 2p $^{2}$P$_{3/2}$ &
1031.912 & $4.59\pm0.16$\\
O {\sc vi} & 2s $^{2}$S$_{1/2}$ -- 2p $^{2}$P$_{1/2}$ &
1037.613 & $2.26\pm0.08$\\
Si {\sc iii} & 3s3p $^{3}$P$_{0}$ -- 3s3d $^{3}$D$_{1}$ &
1108.358& $0.020\pm0.020$\\
Si {\sc iii} & 3s3p $^{3}$P$_{1}$ -- 3s3d $^{3}$D$_{1,2}$ &
1109.96 bl.& $0.065\pm0.020$\\
Si {\sc iii} & 3s3p $^{3}$P$_{2}$ -- 3s3d $^{3}$D$_{1,2,3}$ &
1113.20 bl.& $0.121\pm0.020$\\
S {\sc iv} & 3s$^{2}$3p $^{2}$P$_{1/2}$ -- 3s3p$^{2}$ $^{2}$D$_{3/2}$ &
1062.678& $0.038\pm0.020$\\
S {\sc iv} & 3s$^{2}$3p $^{2}$P$_{3/2}$ -- 3s3p$^{2}$ $^{2}$D$_{5/2}$ &
1072.996& $0.056\pm0.020$\\
S {\sc vi}$^a$ & 3s $^{2}$S$_{1/2}$ -- 3p $^{2}$P$_{3/2}$ &
933.378 & $0.24\pm0.08$ \\ \hline
\end{tabular}

\noindent $^{a}$ Measured from 1-SiC spectra.

\end{table*}

Redfield et al. (2002) have published independent 
measurements of these line fluxes in the $\epsilon$~Eri spectrum. In almost
every case, the fluxes here agree with those given by Redfield et al.
(2002) to within the quoted 
errors. The only exception is the C~{\sc iii} 1175-\AA~
multiplet; in this case, Redfield et al. (2002) report a flux which is 
smaller by $\sim 20$ per cent than that found here, while the errors are 
only $\sim 10$ per cent. 
It is noted that their fit 
(their fig. 10) does appear to underestimate the peak fluxes in this
multiplet.
However, this difference is not significant given the other
sources of uncertainty in the modelling.
The total C~{\sc iii} 1175-\AA~flux measured from STIS (Table 1) and 
that derived from {\it FUSE} (Table 2) are very similar (they agree to 
within the measurement uncertainties). This argues against 
short term variations in the emission line fluxes since the STIS 
and {\it FUSE} data were recorded nine months apart.

\subsection{Interstellar absorption}

Interstellar absorption can be clearly identified in the line profiles of
several of the resonance lines.
This is important since it allows information to be derived about the
local interstellar medium (see, for example, Wood et al. 2000 or
Redfield \& Linsky 2002 for such studies involving spectra of $\epsilon$~Eri)
but also because it must be taken into consideration when analysing the
spectrum of the star.
The largest effect is in lines of the H~{\sc i} Lyman series (this H~{\sc i} 
absorption has been studied by Dring et al. (1997) who used it to determine
the interstellar H~{\sc i} and D~{\sc i} column densities).
However, there is also significant absorption in the Mg~{\sc ii} 
h and k lines and several transitions involving the
ground states of C~{\sc ii} and Si~{\sc ii}. The C~{\sc ii} 
and Si~{\sc ii} lines affected are listed in Table~3 which also gives
estimated line centre opacities ($\tau_0$)
for interstellar absorption in each line.
These have been computed with the C/H abundance for the interstellar medium
given by Cardelli et al. (1996) (gas phase C/H ratio of $1.4 \times 10^{-4}$)
and assuming a solar photospheric value for the relative Si/C abundance
(Grevesse \& Sauval 1998).
The interstellar H~{\sc i} column density was taken
from Dring et al. (1997) and the interstellar line width parameter was
estimated from the observed interstellar absorption in the Mg~{\sc ii} lines.
For the calculations of $\tau_0$, it has been assumed that all the 
interstellar C and Si is
singly ionised -- thus the listed values are upper limits. 

\begin{table}
\caption{Lines of C~{\sc ii} and Si~{\sc ii} affected by interstellar 
absorption and their
calculated interstellar line centre opacities ($\tau_{0}$)
The absolute values of $\tau_{0}$ are
uncertain but the relative
values are reliable (see text).}
\begin{tabular}{|c|c|c|}    \hline
Ion & $\lambda_{0}$ (\AA) & $\tau_{0}$ \\ \hline
C~{\sc ii} & 1334.5 & 5.6 \\
C~{\sc ii} & 1036.3 & 4.4 \\
Si~{\sc ii} & 1190.4 & 1.7 \\
Si~{\sc ii} & 1193.3 & 3.5 \\
Si~{\sc ii} & 1260.4& 7.5 \\
Si~{\sc ii} & 1304.4&0.2\\
Si~{\sc ii} & 1526.7 & 1.0 \\
Si~{\sc ii} & 1808.0 & $2 \times 10^{-2}$ \\ \hline
\end{tabular}
\end{table}

The relative values of $\tau_0$
suggest that the 1260.4-\AA~line should show the greatest
interstellar absorption of the Si~{\sc ii} lines and that there should be
very little effect on the 1526.7, 1304.4 or 1808.0-\AA~lines. 
This is consistent with the observed line profiles -- of the Si~{\sc ii} 
lines, only the 1260.4-\AA~transition shows evidence of self-reversal (in 
addition to line blending).
The C~{\sc ii} resonance lines at 1334.5~\AA~(observed with STIS) and
1036.3~\AA~(observed with {\it FUSE}) are expected to have similar optical
depths due to interstellar absorption and both show evidence of absorption.

\section{Ionisation balance calculations}

Ionization equilibrium populations are needed
for quantitative analyses of stellar atmospheres. Above the chromosphere,
for most ions the dominant ionisation processes are
collisional; either direct collisional ionisation or collisional excitation
followed by autoionisation (Goldberg, Dupree \& Allen 1965).
Photoionisation is usually not important.
Both radiative recombination and
dielectronic recombination (Burgess 1964) must be included.
Currently, the most widely used ionisation 
balance calculations in the literature are those of Arnaud \& Rothenflug
(1985) and Mazzotta et al. (1998) or, for ions of iron, those 
by Arnaud \& Raymond (1992). 
More recently high
precision calculations have been made for specific elements or ions (e.g. 
Nahar 1995 [Si~{\sc i}]; Nahar 1996 [Si~{\sc ii}]; 
Nahar \& Pradhan 1997 [C and N]; Nahar 1999 [O]; Nahar, Pradhan \& Zhang 2000
[C, including relativistic effects]). In common with the earlier work
(Arnaud \& Rothenflug 1985), usually only the
zero-density limit is considered for recombination rates.

Burgess (1965) pointed out that non-zero density reduces the dielectronic
recombination rate because highly excited states effectively become part of 
the continuum. Jordan (1969a) attempted to account for this in an 
approximate way in her ionisation balance calculations and
Judge et al. (1995) used recombination rates
corrected for finite density (based on calculations by Summers 1974a,b)
in their analysis of solar data, but none of the 
more recent calculations mentioned above address this issue. The effects
are 
significant for $\epsilon$~Eri, where the transition region pressure
(and therefore density) is higher than in the Sun -- a reduction of the
dielectronic recombination rate will alter the ionisation balance, affecting,
in particular, the important Li-like and Na-like ions (the 
sensitivity of these ions
is apparent from the calculations of Jordan 1969a).
It is well known from solar work
that emission measures derived from Li-like and Na-like 
ions often appear anomalous 
(Burton et al. 1971, Dupree 1972, Judge et al. 1995). 
It has also been noted in analyses of other cool stars (e.g. Jordan et al.
1987).
Similar results are found for $\epsilon$~Eri;
if the standard Arnaud \& Rothenflug (1985) ionisation balance calculations 
are adopted, 
the loci of Li-like and Na-like ions tend 
to lie too high compared with those of other ions (see Section 4). 
In view of
the sensitivity of these ions to the density-dependent reduction to the
recombination rates it is necessary to attempt to include
this effect in the ionisation balance.

Summers (1972, 1974a,b) computed recombination rates as a function of both 
temperature and electron density ($N_{e}$)
for various elements, including those of primary
interest here (C, N, O, Si). Rates computed  
at electron densities of
$\log N_{e}[\mbox{cm$^{-3}$}]=8.0$ and 12 were taken from
Summers (1974b). For C, N, O and Si these rates were extrapolated
linearly against temperature in $\log$-$\log$ space to find recombination
rates appropriate for the mean transition region pressure in $\epsilon$~Eri
($\log P_{e}[\mbox{cm$^{-3}$~K}] = 15.68$, Jordan et al. 2001a). The 
recombination rates determined in this way for several important ions 
(C~{\sc iv}, N~{\sc v} and Si~{\sc iv}) are different by several hundred
percent from the zero-density limit; this is sufficiently large to 
noticeably affect the ionisation balance.

These extrapolated recombination rates were used together with 
collisional ionisation rates from Bell et al. (1983) to compute the 
fractional ion populations as a function of temperature for C~{\sc ii}
-- {\sc iv}, N~{\sc iii} -- {\sc v}, O~{\sc iii} -- O~{\sc vi} and
Si~{\sc ii} -- {\sc iv}. Photoionisation has been neglected for
these calculations, as has charge exchange for the C, N and O ions. Charge
exchange (with rates taken from Arnaud \& Rothenflug 1985) was included
for Si. 
As expected, the ionisation fractions computed differ most significantly
from those of Arnaud \& Rothenflug (1985) for Li-like and Na-like ion, but
there are also differences for several lower ionisation stages.
The reduction in recombination rates
causes the peak ionisation fraction to occur at lower temperatures
for many ions (as apparent in Jordan 1969a).

These effects are illustrated in Fig. 1 which shows ionisation balance 
calculations for C. Two sets of zero-density calculations are shown
(Arnaud \& Rothenflug 1985; Nahar \& Pradhan 1997) along with 
Jordan's (1969a) results for a solar pressure and the new computations
for $\epsilon$~Eri. Generally, the zero-density calculations 
(Arnaud \& Rothenflug 1985 and  Nahar \& Pradhan 1997) agree well with
each other but there are significant differences from the finite density
cases. The agreement between the new calculations and those of
Jordan (1969a) is good -- although Jordan's calculations were intended for 
densities in the quiet Sun, she found later that the density sensitivity
was overestimated and that
the results are more applicable to higher pressures
such as those found in active regions, or $\epsilon$~Eri. 
The results for the other elements considered (N, O, Si)
are all similar. The largest differences between the new calculations
and those by Jordan (1969a) are for Si since the latter did not include
charge exchange processes.

\begin{figure*}
\epsfig{file=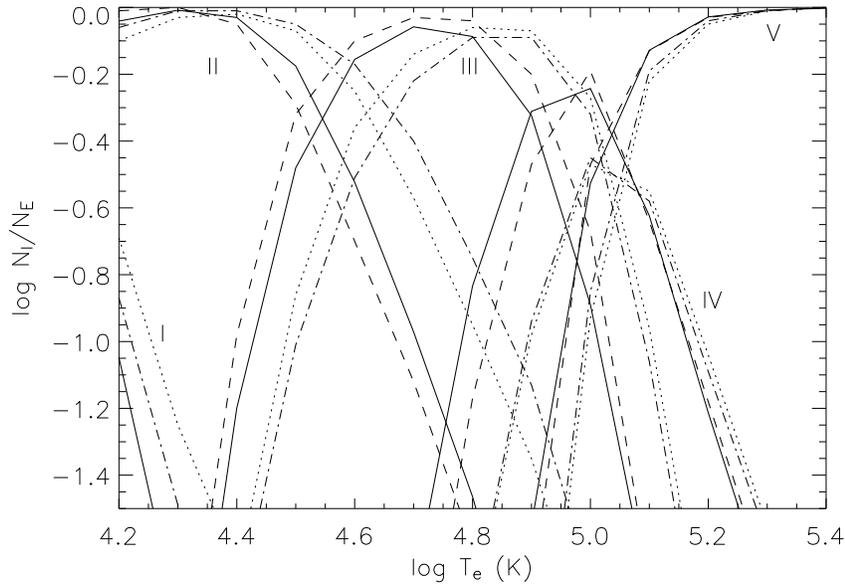, width=12cm}
\caption{Ionisation fractions for C ($N_{\mbox{I}}/N_{\mbox{E}}$) 
as a function of temperature. C~{\sc i} -- C~{\sc v} are shown, although
only results for C~{\sc ii} -- C~{\sc iv} are needed in Section 4. The
curves are labelled with the ionisation stage they represent. The solid
curves are the ionisation fractions computed here. The dashed curves
are the finite density calculations of Jordan (1969a); the dotted
curves are the Nahar \& Pradhan (1997) ion fractions and the dot-dash
curves are from Arnaud \& Rothenflug (1985).}
\end{figure*}

The new ionisation balance calculations for C, N, O and Si are used in the
emission measure modelling (see Section 4) and the 
recombination rates computed for finite density are employed in the 
atomic models of Si and C discussed in Section 5.2. 
The computed ionisation fractions are tabulated in Appendix~A. 
The accuracy of these ion fractions
is limited by the accuracy of the Summers (1972,1974a,b) 
recombination rates; there is now a real need for modern calculations
of di-electronic recombination rates at finite density.

\section{Emission measure distribution (EMD)}

\subsection{Emission measure loci}
In this section the fluxes presented in Section 2 are
used to construct loci which constrain the mean emission measure distribution
$Em^{0.3}(T_{e})$. The principles of this method have been discussed at length
elsewhere (e.g. Jordan \& Brown 1981). The notation and definitions 
used by Sim \& Jordan (2003b) are adopted here. 

The standard method of constructing emission measure loci from line fluxes
is only valid for optically thin lines. For optically thick lines
radiative transfer calculations are required. 
In order to construct a
``first-cut'' model, however, even optically thick lines can be included in
the emission measure analysis by assuming that
they are part of a multiplet which is 
effectively optically thin. For lines in which scattering is
the only opacity source this assumption should not lead to errors of
more than a factor of two. This approach is used in this section. 
In Section 5, full radiative transfer calculations
are discussed and used to improve the calculations presented here.

Throughout the following discussion, solar photospheric 
values have been adopted for the
relative abundances of the elements
(from Grevesse \& Sauval 1998), with absolute values scaled to the
$\epsilon$~Eri metallicity derived by Drake \& Smith (1993).
The observed line fluxes have been converted to fluxes at the stellar surface
using the stellar parameters adopted by Jordan et al. (2001b, table 1).
We investigate later the differences which result from adopting the
recent solar abundances derived by Asplund, Grevesse \& Sauval (2004).

\subsection{Density-insensitive STIS lines}

For many of the strong emission lines in the STIS wavelength range, the
contribution function $K(T_{e})$\footnote{An explicit 
expression for the contribution function ($K(T_{e})$) in 
terms of abundance, ionisation fraction, degree of excitation, electron
number density and Einstein-A coefficient is given by Sim \& Jordan (2003b).}
 is almost completely insensitive to the
density. This generally occurs for permitted transition in which the
upper level is predominantly
excited by collisions from the ground state. Such lines
place the strongest constraints on the EMD since their
loci are independent of any uncertainty in the electron pressure.

Table~4 lists the density-insensitive lines/multiplets from the STIS lines
(the insensitivity criterion was taken to be a variation in $K(T_{e})$ of
less than ten per cent resulting from a variation of $\log P_{e}$ by 
$\pm 0.3$~dex around the mean transition region pressure).
Table~4 also indicates whether the lines 
may reasonably be assumed to be
optically thin or whether only the multiplet as a whole is likely to be thin
(it is stressed that optically thick lines are treated with 
radiative transfer calculations at a later stage, see Section 5).

\begin{table}
\caption{Density-insensitive lines used for emission measure modelling.
Where an entry is given in parenthesis it is likely that
the lines are affected by radiative transfer effects which are dealt with
in Section 5.}
\begin{tabular}{lcc} \hline
Ion & $\lambda$ (\AA) &  Assumed opacity \\ \hline
(C~{\sc ii} & 1335 & multiplet effectively thin) \\
C~{\sc iii} & 1175 & thin \\
C~{\sc iv} & 1550 & thin \\
N~{\sc v} & 1240 & thin \\
O~{\sc v} &  1371 & thin \\
(Mg~{\sc ii} & 2800 & multiplet effectively thin) \\
Al~{\sc iii} & 1854 & effectively thin\\
Si~{\sc ii} & 1810 & multiplet effectively thin \\
Si~{\sc ii} & 1530 & multiplet effectively thin \\
Si~{\sc ii} & 1308 & multiplet effectively thin \\
Si~{\sc ii} & 1260 & multiplet effectively thin \\
Si~{\sc ii} & 1195 & multiplet effectively thin \\
Si~{\sc iii} & 1206 & effectively thin \\
Si~{\sc iv} & 1400 & thin \\
S~{\sc ii} & 1255& multiplet effectively thin \\
S~{\sc iv} & 1406&  thin \\ \hline
\end{tabular}
\end{table}

The multiplets listed in Table 4 have been used to construct emission measure 
loci. The $K(T_{e})$ contribution functions were computed
with the CHIANTI database (v3.03; Dere at al. 1997; Dere et al. 2001).
For C, N, O and Si the ionisation fractions were computed as described in 
Section 3. For other elements the Arnaud \& Rothenflug (1985) ionisation 
fractions were adopted. The loci computed are plotted in Fig. 2.

\begin{figure*}
\epsfig{file=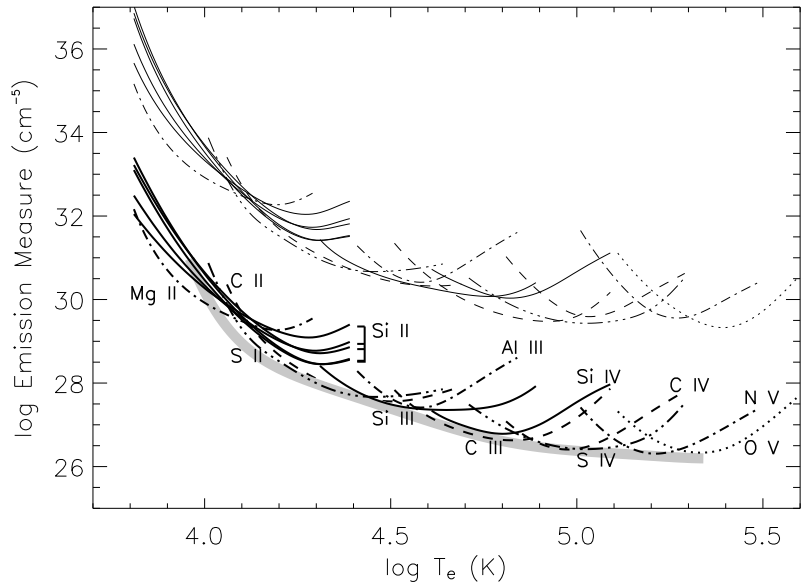, width=12cm}
\caption{Emission measure loci from density-insensitive lines in the 
$\epsilon$~Eri STIS spectrum. The solid curves are for lines of Si; the
dashed curves for C; the dotted curve for O; dot-dash curves are used for
N, Al and Mg; and the triple-dot-dash curves are for S. The heavy lines show 
the loci computed using the new ionisation fractions from Section 3
and are labelled with the ion they represent. The light
lines show the loci computed if all ion fractions are taken from
Arnaud \& Rothenflug. These curves have been off-set by +3~dex, for 
clarity. The broad, light curve is the mean EMD corresponding
to DEMD-A (see Section  4.5).}
\end{figure*}

The loci suggest a smooth EMD decreasing from
the chromosphere to a minimum somewhere in the mid-transition region
at around
$\log T_{e} \approx 5.3$. There is a sufficiently large number 
of lines to provide temperature coverage throughout this
part of the atmosphere, placing constraints on the EMD
that are at least as complete as those for the
Sun.

The lightly drawn loci in Fig. 2 (offset by +3~dex for clarity)
show the constraints derived
from the same line fluxes if the Arnaud \& Rothenflug (1985) ionisation
calculations are used for every ion. In this case, C~{\sc iv} and 
Si~{\sc iv} are clearly anomalous by factors $>2$ and N~{\sc v} also appears 
to lie rather high. However,
when the new ionisation calculations are adopted (Section 3) the loci 
from these Li-like and Na-like ions are in much better agreement with
those of other ions. The agreement is still not perfect, in
particular the Si~{\sc iv} locus still lies higher than that of C~{\sc iii} by
about 30 per cent, but this difference is small compared to the likely
residual uncertainty in the ionisation balance calculations.
The Al~{\sc iii}
locus is also rather high; although this may be the result of blending
in the observed Al~{\sc iii} line, it may also be the result of errors
in the ionisation balance; like C~{\sc iv}, Al~{\sc iii} is an alkali but no
new ionisation calculations have been performed for this ion.

It is concluded that all the loci which appear significantly
anomalous when 
the Arnaud \& Rothenflug (1985) ionisation balance calculations are adopted
are likely to be the result of errors in the recombination rates used in that
ionisation balance, owing to the density dependence of dielectronic
recombination. 
This is in contrast to the analysis of solar data by Judge et al. (1995);
they
find that loci derived for alkali-like ions are still discordant, even if
they account for the density-sensitivity of dielectronic recombination
and they interpret this as evidence of dynamic or diffusive effects
(although they do state that uncertainties in the recombination rates may 
still be significant).
\footnote{However,
Judge et al. (1995) used a mixture of lines
above and below the H~Lyman edge at 911~\AA, whereas all the 
lines used here
lie above 911~\AA.
Judge et al. (1995) rejected absorption by the Lyman continuum as a
source of the discrepancies that they found, partly on the basis that 
centre-to-limb differences have not been observed. 
However, Burton et al. (1973) established
the magnitude of this by observing the limb-to-disc ratio of pairs of 
optically thin lines
from a given ion, where one was found at $\lambda > 911$~\AA~and the other
at $\lambda < 911$~\AA. 
This amounts to a factor of up to $\sim$~3
between $\lambda \sim 765$ and 835~\AA. 
These ratios
are independent of any atomic physics, abundances or intensity calibration.
The effect expected for stellar observations should, of course,
be less than this.}
This issue cannot be fully resolved without reliable recombination rates.

\subsection{Density-sensitive STIS lines}

The density-insensitive {\it STIS} loci can be supplemented
by loci derived from density-sensitive lines/multiplets. Table 5 lists the
density-sensitive multiplets from the STIS data. Most of these multiplets
and the associated atomic data
have been discussed by Jordan et al. (2001a) where they were used to deduce
the mean transition region pressure. The Si~{\sc ii} 2334-\AA~multiplet
is not used here; this multiplet is density-sensitive but there are line
blends and obvious problems with the atomic data.

\begin{table}
\caption{Density-sensitive lines used for emission measure modelling.}
\begin{tabular}{lcc} \hline
Ion & $\lambda$ (\AA) & Opacity \\ \hline
N~{\sc iv} & 1486 & thin \\
O~{\sc iii} & 1663 & thin \\
O~{\sc iv} & 1400 & thin \\
O~{\sc v} & 1218 & thin \\
Si~{\sc iii} & 1892 & thin \\
Si~{\sc iii} & 1300 & thin \\ \hline
\end{tabular}
\end{table}

\begin{figure*}
\epsfig{file=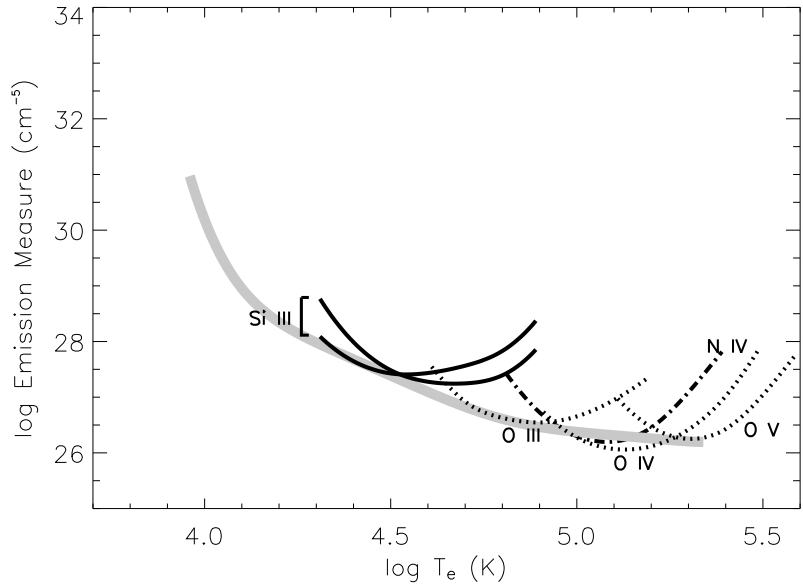, width=12cm}
\caption{Emission measure loci from density-sensitive lines in the 
$\epsilon$~Eri STIS spectrum. The solid curves are for lines of Si; the
dotted curve for O; the dot-dash curve for
N. The light broad curve shows the same EMD plotted in Fig. 2.}
\end{figure*}

The loci deduced from the density-sensitive loci are plotted in Fig 3. 
In general there is good agreement between the density-sensitive and
density-insensitive loci when 
$\log P_{e}[\mbox{cm$^{-3}$~K}] = 15.68$ is adopted. 
The N~{\sc iv} and, in particular, the O~{\sc iv} 
loci lie lower than those from density-insensitive lines at similar 
temperatures. 
A similar effect has been noted in the Sun; Judge \& Brekke (1994) 
found that loci derived from density-sensitive lines lie lower than those
from resonance lines by a significant factor.
In part this may be due to uncertainties in the atomic
data for these transitions (it is known that there are inadequacies in 
the O~{\sc iv} atomic data -- see the discussion by Jordan et al. 2001a). 
There are also complications in the interpretation of the 
loci from density-sensitive multiplets resulting from inhomogeneity in the
atmosphere (Judge \& Brekke 1994). 
If the atmosphere is inhomogeneous then loci 
derived from density-insensitive lines will represent unbiased
averages over the inhomogeneous structure.
In contrast, the density-sensitive lines will 
form predominantly in regions with higher or lower density than average 
resulting in a pressure-weighted average. 
To investigate this
effect it is necessary to quantify the nature of the sensitivity of
the density-sensitive lines. 
Fig. 4 shows loci derived for the density-sensitive lines 
at different pressures.
For most of the lines, a higher pressure means that less flux is formed
for a given emission measure (i.e. the emission measure locus lies higher 
for a given flux) and vice versa. The Si~{\sc iii} 1300-\AA~multiplet is
an exception because of the influence of the density on the
metastable $^3$P populations.
The N~{\sc iv}, O~{\sc iii} and Si~{\sc iii} 1892-\AA~loci are the most
sensitive to errors in the pressure, changing by about 50 per cent when
the pressure is varied by a factor of two. O~{\sc iv} and O~{\sc v} are
interesting because, although less sensitive than O~{\sc iii}, they respond
more to a high density than a low density. Thus if there is a 
range of densities present in the atmosphere
those places with density lower than average will not contribute
significantly more than average, 
but those regions with high density will contribute
significantly less. 
This may partially explain why the O~{\sc iv} locus appears to
lie too low.

\begin{figure*}
\epsfig{file=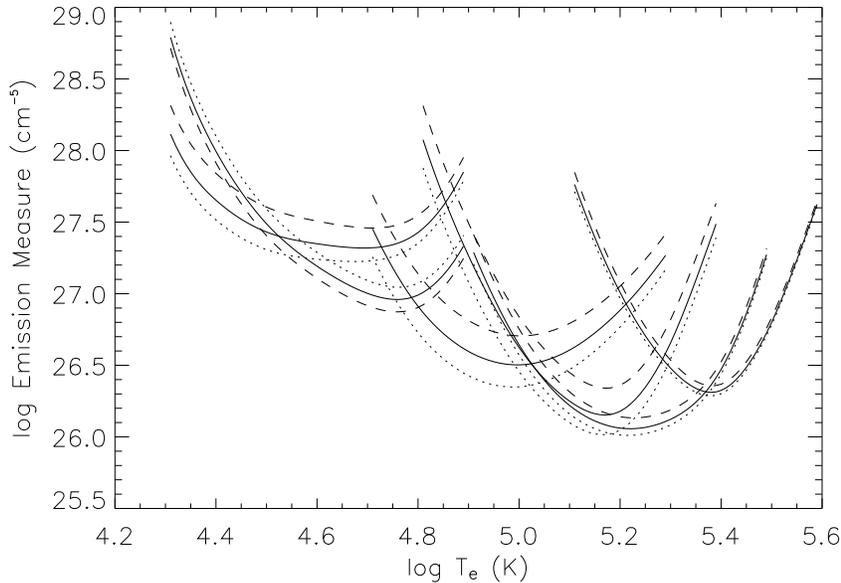, width=12cm}
\caption{Emission measure loci from density-sensitive lines in the
$\epsilon$~Eri {\it STIS} spectrum. The curves are for the same
ions as indicated in Fig 3. The solid curves were
calculated using the best mean electron pressure, the
dotted curves using a pressure smaller than the mean by a factor of two
and the dashed curves using a pressure higher than the mean by a factor of 
two. Note that the scale is significantly expanded compared to 
Figs 2 and 3.}
\end{figure*}

\subsection{{\it FUSE} lines}

The multiplets measured in the {\it FUSE} spectra provide useful supplementary
constraints. Table 6 lists the multiplets observed with {\it FUSE} that have
been used to construct emission measure loci. The loci are shown in
Fig. 5.

\begin{table}
\caption{{\it FUSE} lines used for emission measure modelling. The last
column gives the {\it FUSE} channel in which the lines are observed.
Only the pointing of the 1-LiF is accurate. 
The other channels may lose photons due to channel drift 
making the measurements from these channels prone to large errors.}
\begin{tabular}{lccc} \hline
Ion & $\lambda$ (\AA) & Opacity& Channel \\ \hline
C~{\sc ii} & 1010 & effectively thin& 1A-LiF\\
C~{\sc ii} & 1037 & (effectively thin)& 1A-LiF\\
C~{\sc iii} & 1175 & thin& 1B-LiF\\
C~{\sc iii} & 977 & effectively thin& 1B-SiC\\
N~{\sc iii} & 991 & thin&1A-LiF \\
O~{\sc vi} & 1035 & thin& 1A-LiF\\
Si~{\sc iii} & 1110 & thin& 1B-LiF\\
S~{\sc iv} & 1070 & thin& 1A-LiF\\ 
S~{\sc vi} & 933 & thin& 1B-SiC\\ \hline
\end{tabular}
\end{table}

\begin{figure*}
\epsfig{file=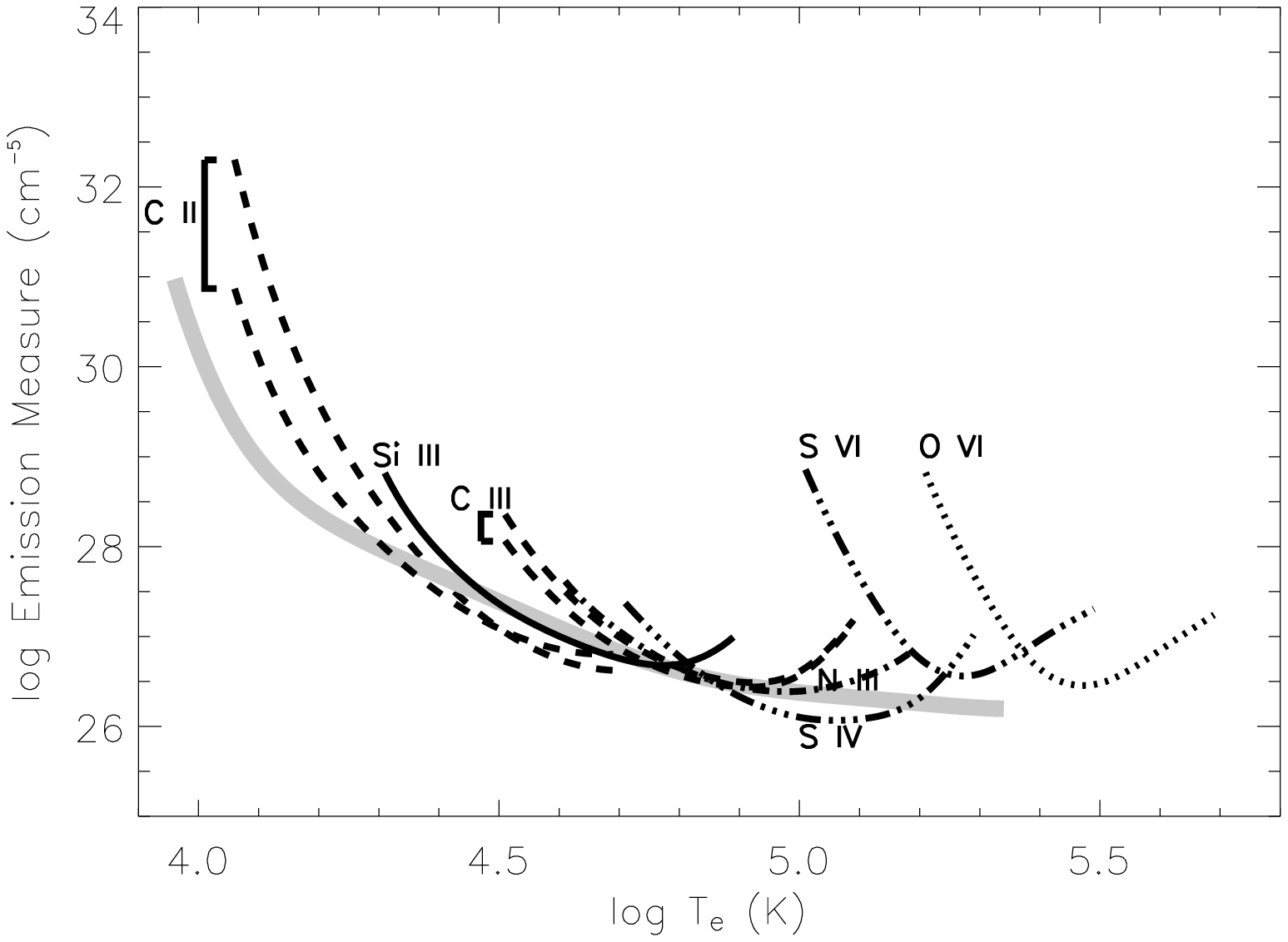, width=12cm}
\caption{Emission measure loci from lines in the 
$\epsilon$~Eri {\it FUSE} spectrum. The solid curve is for the Si multiplet; 
the dashed curves for C; the
dotted curve for O; the dot-dash curve for
N; the triplet-dot-dash curves for S. 
The light, broad curve shows the mean EMD from Fig. 2 for comparison.}
\end{figure*}

Most of the {\it FUSE} loci agree well with each other, and with the 
{STIS} loci. The {\it FUSE} S~{\sc iv} locus is not consistent with the
{STIS} loci. The S~{\sc iv} measurement error is large ($\sim 40$ per cent)
but not sufficient to explain the discrepancy. 
The most probably explanation is a systematic error in the
S~{\sc iv} atomic data.

\subsection{Preliminary mean EMD}

The procedure of constructing emission measure distributions from emission 
measure loci has come under close scrutiny in recent years (e.g.
Judge, Hubeny \& Brown 1997; McIntosh, Brown \& Judge 1998).
Since the EMD is a continuous function but we have only a finite number of
constraints, each with 
an associated uncertainty, it is not possible to determine a unique 
EMD from the 
data alone. Thus the EMD obtained will depend on how it is 
obtained. However, as stressed by Judge et al. (1997), the EMD
determined to be consistent with the data is still important since it
represents the most information that can be extracted from the data alone,
and the range of distributions that are consistent with the data is an
indication of the completeness of the information provided by the data. 

Here, the mean EMD ($Em^{0.3}(T_{e})$) has been constructed by first
obtaining the {\it differential emission measure distribution} (DEMD)
defined by

\begin{equation}
\frac{\mbox{d}Em}{\mbox{d}\log T_{e}} = G(r) f(r) \frac{A(r)}{A_{*}(r)}
N_{e}(T_{e}) N_{H}(T_{e}) \frac{\mbox{d}r}{\mbox{d}\log T_{e}}
\end{equation}
(see Sim \& Jordan 2003b for definitions of the quantities in this
and the following equation). 
A first-cut DEMD was used to compute
fluxes ($F_{*}$) for each of observed multiplets using the integral

\begin{equation}
F_{*} = \int K(T_{e}) \frac{\mbox{d}Em}{\mbox{d}\log T_{e}} \;
\mbox{d}\log T_{e} \; \; \; .
\end{equation}
These were compared with the observed fluxes. The DEMD was then 
modified iteratively until reasonable agreement ($\sim 50$ per cent)
was obtained
between the fluxes predicted by the DEMD and the observed fluxes.
Greatest weighting was given to the density-insensitive lines. 
During this process, it was assumed that the DEMD remains smooth on 
scales $\Delta \log T_{e}[\mbox{K}] < 0.1$~dex, which is comparable to
the typical scale of overlap between adjacent emission measure loci
from the {\it STIS} data.

The DEMD was only specified over the temperature range
$3.8 < \log T_{e} [\mbox{K}] < 5.5$ since outside this range there are
insufficient lines to provide robust constraints. This means that lines
that form partially outside this temperature range (in particular those
of O~{\sc vi} and O~{\sc v}) will be underpredicted by the DEMD.
The 
resulting DEMD (DEMD-A) was used to compute a mean EMD (EMD-A) which
is overplotted in Figs. 2, 3 and 5. The fluxes computed from DEMD-A
using equation (2) are compared with observed fluxes in Table 7.

\begin{table}
\caption{Comparison of calculated fluxes ($F_{c}$) and observed fluxes ($F_{o}$)
for emission measure distribution DEMD-A. The error column gives the 
percentage error on the measured flux.
The top part of the table gives the density-insensitive {\it STIS} lines,
 the middle part gives the density-sensitive {\it STIS} lines and the lower
part gives the {\it FUSE} lines.}
\begin{tabular}{lccc} \hline
Ion & $\lambda$ (\AA)& \% error & {$(F_{c} - F_{o})/F_{o} \times 100\%$} \\ \hline
C~{\sc ii} & 1335 & 3 & -7\\
C~{\sc iii} & 1175 & 10 &13\\
C~{\sc iv} & 1550 & 2&-19\\
N~{\sc v} & 1240 &4&-29\\
O~{\sc v} & 1371 &5&-40\\
Mg~{\sc ii} & 2800 &2&-41\\
Al~{\sc iii} & 1854 &12&-60\\
Si~{\sc ii} & 1810 &1& -55\\
Si~{\sc ii} & 1530&1&-68\\
Si~{\sc ii} & 1308&5&-80\\
Si~{\sc ii} & 1260&13&-67\\
Si~{\sc ii} &1195&30&-71\\
Si~{\sc iii} &1206&3& -17\\
Si~{\sc iv} &1400&3&-29\\
S~{\sc ii} &1255&10&-4\\
S~{\sc iv} & 1406&20&9\\
\\
N~{\sc iv} & 1486&33&30\\
O~{\sc iii} & 1663&17&16\\
O~{\sc iv} &1400&6&89\\
O~{\sc v} &1218&6&-22\\
Si~{\sc iii} &1892&4&5\\
Si~{\sc iii} &1300&7&-22\\ 
\\
C~{\sc ii} & 1010 &25 &30\\
C~{\sc ii} & 1037 &6&56\\
C~{\sc iii} & 977 &2&52\\
C~{\sc iii} & 1175 &2&11\\
N~{\sc iii} & 991 & 24&35\\
O~{\sc vi}& 1035 &4&-50\\
Si~{\sc iii} & 1110 & 30&12\\
S~{\sc iv} & 1070 &43&125\\
S~{\sc vi} & 933 &33&-68\\ \hline
\end{tabular}
\end{table}

DEMD-A predicts fluxes for most of the high-temperature, optically thin lines
(C~{\sc iii}, C~{\sc iv}, N~{\sc v}, Si~{\sc iii}, Si~{\sc iv}, S~{\sc iv},
N~{\sc iv}, O~{\sc iii}) that are in agreement with the observations to
within
$\sim$ 30 per cent, which is comparable to the likely measurement errors for 
the weak lines, and the uncertainty in the ionisation balance
for the alkali-like ions. 
Among the high-temperature lines,
the poorest agreement is for O~{\sc iv} 1400~\AA~ and S~{\sc iv} 1070~\AA.
Both of these multiplets have already been identified as anomalous from the
EMD loci discussed above.
Note that for both Al~{\sc iii} and S~{\sc vi} the underprediction is 
probably the result of errors in the ionisation balance -- these are
both alkali-like ions for which the finite density has not been accounted for
in the recombination rates (see Section 3).

Among the lower-temperature ions (C~{\sc ii}, Mg~{\sc ii} and Si~{\sc ii})
the agreement is less good. However, these multiplets are likely to be
affected by opacity (making equation 2 invalid)
-- their discussion is postponed to Section 5 where
full radiative transfer calculations are presented.

\section{The preliminary atmospheric model}

\subsection{Construction of the model}

In this and subsequent sections, 
EMDs are used to construct new atmospheric models for
$\epsilon$~Eri. These models are used to perform radiative
transfer calculations for the important optically thick lines, the results
of which are then discussed.

The atmospheric models presented are all constructed in one-dimension 
under the
assumption of hydrostatic equilibrium, including both thermal and 
microturbulent contributions to the pressure. Full details of the method
of constructing atmospheric models from EMDs are presented elsewhere
(see e.g. Griffiths \& Jordan 1998 and Harper 1992 for all the relevant
background or Sim 2002 for a complete discussion of the iteration scheme
used here). The Thatcher et al. (1991) model is used
to describe the atmosphere below the temperature range to which the
DEMD extends ($\log T_{e} [\mbox{K}] < 3.8$).

An important ingredient in the modelling is the microturbulent velocity in 
the chromosphere/transition region. Sim \& Jordan (2003a) 
used line width measurements from the $\epsilon$~Eri ultraviolet data
to determine the most-probable 
non-thermal speed ($\xi(T_{e})$) in the upper atmosphere. For the models
presented here, these results have been fitted with the simple form

\begin{equation}
\xi(T_{e}) = \left\{ {\begin{array}{cl}
			-31.56 + 10.716 \log T_{e} & 3.8 < \log {T_{e}} < 4.9 \\
			20.95 & \log {T_{e}} > 4.9 \\
			\end{array}} \right\} \mbox{km~s}^{-1}
\end{equation}
which is an adequate representation of the line-width measurements.
Initially, it is assumed that {\it all} the observed non-thermal
motion can be classified as microturbulence (i.e. $\xi(T_{e})_
{\mbox{\scriptsize micro}} = \xi(T_{e})$). Later, a small amount of
macroturbulence is introduced (Section 6).
For $\log T_{e}[\mbox{K}] < 3.8$, the microturbulence is extrapolated linearly
in $\log T_{e}$ to agree with the measured photospheric microturbulent
velocity (1.25 km~s$^{-1}$; Drake \& Smith 1993) at the 
temperature minimum.

\begin{table*}
\caption{Fluxes computed using MULTI ($F_{c}$) compared with observed fluxes
($F_{o}$) for lines of H, Mg, Si and C. $F_{c}$ is given in units of
$10^{-13}$~ergs~cm$^{-2}$~s$^{-1}$. The first section gives the density-
insensitive lines from the {\it STIS} spectrum, the second gives the
density-sensitive lines from {\it STIS} and the third gives the lines from 
{\it FUSE}. Computed fluxes are given for models A, B, C and D. Following 
the model B fluxes,
values are given for computations from a model identical to model B except
 that the
Asplund et al. (2004) relative abundances are adopted.}
\begin{tabular}{lccccccccccc}\hline
Ion & $\lambda_{0}$ & \multicolumn{4}{c}{$F_{c}$} 
&\multicolumn{4}{c}{$(F_{c} - F_{o})/F_{o} \times 100\%$}\\
& (\AA) & Model A & Model B & Model C &Model D& Model A & Model B & Model C &Model D\\ \hline
H~{\sc i}$^{a}$ & 1215.671 & 634. & 441 (410) & 302 & 220 &124 &176 (156) & 88.8 & 37.5\\
C~{\sc ii}$^{a}$ & 1334.532& 3.30 & 3.41 (2.49) & 1.63 & 3.25 & 75.5 &81.4 (32.4) & -13.3 & 72.9\\
C~{\sc ii} & 1335.68 bl.& 8.29 & 8.68 (6.26) & 4.04 & 8.81 & 71.3 &79.3 (29.3) &-16.5 & 82.0\\
C~{\sc iii}& 1174.933& 0.541 &0.556 (0.404)& 0.453 & 0.589&32.0 &35.6 (-1.5)&10.5& 43.7\\
C~{\sc iii}& 1175.263& 0.425 &0.437 (0.318)& 0.357 & 0.441&3.7 &6.6 (-22.4) &-12.9& 7.6\\
C~{\sc iii}& 1175.69 bl.& 1.908 &1.961 (1.429) & 1.602 & 1.976&28.1 &31.6 (-4.1) &7.5& 32.6\\
C~{\sc iii}& 1175.987& 0.421 &0.433 (0.315) & 0.353 & 0.438&2.7 &5.6 (-23.2) &-13.9& 6.8\\
C~{\sc iii}& 1176.370& 0.528 &0.543 (0.395) & 0.444 & 0.545&14.8&18.0 (-14.1) &-3.5&18.5\\
C~{\sc iv}& 1548.187& 3.83 &4.39 (3.19) & 4.32 & 4.36&-31.5&-21.5 (-42.9) &-22.7& -22.0\\
C~{\sc iv}& 1550.772& 1.92 &2.20 (1.59) & 2.16 & 2.18&-31.9&-22.0 (-43.6) &-23.4&-22.7\\
Mg~{\sc ii}$^{a}$&2796.352& 314. &348 (312) & 351 & 213 &3.0& 14.1 (2.3) &15.1&-30.2\\
Mg~{\sc ii}$^{a}$&2803.531& 196 &218 (193) & 228 & 138 &-15.5& -6.0 (-16.8)&-1.7&-40.5\\
Si~{\sc ii}&1190.416& 0.031&0.060 (0.055)& 0.074 & 0.056&-69.0&-40.0 (-45.0) &-26.0&-44.0\\
Si~{\sc ii}&1193.290& 0.033&0.057 (0.054) & 0.075 & 0.043&-63.3&-36.7 (-40.0) &-16.7&-52.2\\
Si~{\sc ii}&1194.500& 0.092&0.131 (0.122)& 0.170 & 0.101&-48.9&-27.2 (-32.2)&-5.6&-43.9\\
Si~{\sc ii}&1197.394& 0.013&0.034 (0.033)& 0.045 & 0.028&-87.0&-66.0 (-67.0)&-55.0&-72.0\\
Si~{\sc ii}$^{a}$&1260.422& 0.154&0.215 (0.192) & 0.262 & 0.183&-19.4&12.6 (0.5) &37.2&-4.2\\
Si~{\sc ii}&1264.738& 0.305&0.450 (0.400)& 0.530 & 0.404&-39.6&-10.9 (-20.8) &5.0&-20.0\\
Si~{\sc ii}&1265.002& 0.065&0.107 (0.094)& 0.119 & 0.107&-62.2&-37.8 (-45.3) &-30.8&-37.8\\
Si~{\sc ii}&1304.370& 0.047&0.075 (0.066)& 0.085 & 0.071&-64.7&-43.6 (-50.4)&-36.1&-46.6\\
Si~{\sc ii}&1309.276& 0.071&0.107 (0.095)& 0.126 & 0.096&-68.0&-51.8 (-57.2)&-43.2&-56.8\\
Si~{\sc ii}&1526.707& 0.294&0.570 (0.508)& 0.661 & 0.478&-42.1&12.2 (0.0) &30.1& -5.9\\
Si~{\sc ii}&1533.431& 0.405&0.764 (0.685)& 0.933 & 0.604&-33.1&25.9 (12.9) &53.7&-0.5\\
Si~{\sc ii}&1808.013& 3.28&5.90 (5.23)& 5.69 & 5.31&-48.7&-7.7 (-18.2) &-11.0&-16.9\\
Si~{\sc ii}&1816.928& 6.49&11.26 (10.00)& 10.76 & 9.55 &-37.7&8.1 (-4.0) &3.3&-8.3\\
Si~{\sc ii}&1817.451& 1.57& 2.61 (2.29)& 3.18 & 2.67 &-50.9&-18.4 (-28.4) &-0.6&-16.6\\
Si~{\sc iii}&1206.500& 3.34&3.47 (3.09) & 4.05 & 4.05 &-7.2&-3.6 (-14.2) &12.5& 12.5\\
Si~{\sc iv}&1393.755& 1.53& 1.561 (1.396)& 3.052 & 1.556&-33.5&-32.1 (-39.3) &32.7& -32.3\\
Si~{\sc iv}&1402.770& 0.772& 0.787 (0.704)& 1.538 & 0.783&-36.2&-35.0 (-41.8) &27.1&-35.3\\ 
\\
Si~{\sc iii} & 1294.545& 0.038 & 0.038 (0.034)& 0.047 & 0.039&-26.9&-26.9 (-34.6) &-9.6& -25.0\\
Si~{\sc iii} & 1296.726& 0.030&0.030 (0.027)& 0.037 & 0.031&-26.8&-26.8 (-34.1) &-9.8& -24.4\\
Si~{\sc iii} & 1298.93 bl.& 0.134&0.135 (0.121)& 0.166 & 0.138&-19.8&-19.2 (-27.5) &-0.6&-17.4\\
Si~{\sc iii} &1301.149& 0.025&0.025 (0.023)& 0.029 & 0.025&-10.7&-10.7 (-17.9) &3.6&-10.7\\
Si~{\sc iii} &1303.323& 0.037&0.037 (0.034)& 0.046 & 0.038&-40.3&-40.3 (-45.2) &-25.8&-38.7\\
Si~{\sc iii} &1892.030& 1.25&1.19 (1.07)& 1.37 & 1.30 &0.0&-4.8 (-14.4)& 9.6 & 4.0\\
\\
H~{\sc i}$^{a}$ & 972.537  & 2.81& 2.80 (2.73) & 2.02 & 1.05&329 &344 (333) & 221 & 66.7\\
H~{\sc i}$^{a}$ & 1025.722 & 10.46&11.02 (10.69) & 7.74 & 4.84&172&201 (192) & 111 & 32.2\\
C~{\sc ii} & 1009.858& 0.018&0.019 (0.013) & 0.009 & 0.020&50.0 &58.3 (8.3) & -25.0 & 66.7\\
C~{\sc ii} & 1010.083& 0.037&0.037 (0.027) & 0.018 & 0.039&37.0&37.0 (0.0) & -33.3 &44.4\\
C~{\sc ii} & 1010.371& 0.055&0.055 (0.040) & 0.027 & 0.058&19.6&19.6 (-13.0)& -41.3 &26.1\\
C~{\sc ii}$^{a}$ & 1036.337& 0.448&0.403 (0.289)& 0.192 & 0.420&273&236 (141) & 60.0 & 250\\
C~{\sc ii} & 1037.018& 0.681 &0.627 (0.454) & 0.304 & 0.611&134&116 (56.0) & 4.5 &110\\
C~{\sc iii} & 977.020& 8.85&9.09 (6.45) & 6.76 & 9.99&89.5&94.6 (38.1) & 44.8 &114\\
C~{\sc iii}& 1174.933& 0.541&0.556 (0.404) & 0.453 & 0.589&24.4&27.8 (-7.1) & 4.1 &35.4\\
C~{\sc iii}& 1175.263& 0.425&0.437 (0.318) & 0.357 & 0.441&0.0&2.8 (-25.2) & -16.0 &3.8\\
C~{\sc iii}& 1175.69 bl.& 1.908&1.961 (1.429) & 1.602 & 1.976&25.6&29.1 (-5.9) & 5.5 &30.1\\
C~{\sc iii}& 1175.987& 0.421&0.433 (0.315)& 0.353 & 0.438&4.0&6.9 (-22.2)& -12.8&8.1\\
C~{\sc iii}& 1176.370& 0.528&0.543 (0.395)& 0.444 & 0.545&13.1&16.3 (-0.15) &-4.9&16.7\\
Si~{\sc iii}&1108.358& 0.023&0.023 (0.021)& 0.030 & 0.023&15.0&15.0 (5.0) & 50.0&15.0\\
Si~{\sc iii}&1109.96 bl.& 0.065&0.066 (0.059)& 0.085 & 0.067&0.0&1.5 (-9.2) &30.8&3.1\\
Si~{\sc iii}&1113.20 bl.& 0.131&0.133 (0.119)& 0.174 & 0.136&8.3&9.9 (-1.7) &43.8&12.4\\ \hline
\end{tabular}

{$^{a}$ Line affected by interstellar absorption.}

\end{table*}

\subsection{Radiative transfer calculations}

Radiative transfer calculations
were performed with the MULTI code (Scharmer \& Carlsson
1985a,b; Carlsson 1986). In addition to the atmospheric models, as input
MULTI requires atomic models for the ions that give rise to the
lines under consideration. Brief details of the 
adopted atomic models are given below.

\subsubsection{Hydrogen}

The H model used contains levels $n=1$ to $n=8$ and the continuum. The
radiative transfer calculations for H make use of the 
Hubeny \& Lites (1995) modifications to MULTI which allow the effects of
Partial Redistribution (PRD) to be taken into account for Lyman $\alpha$ and
$\beta$. (See Hubeny \& Lites 1995 and Sim 2001 for discussions of PRD
effects in cool star atmospheres.)

\subsubsection{Magnesium}

A ten level Mg model, developed by P. Judge, was used. The model describes the
first three ionisation stages of Mg. PRD is included in the h and k-lines
of Mg~{\sc ii} using the Uitenbroek (1989) modifications to MULTI.

\subsubsection{Silicon}

In view of the large number of Si lines observed in the {\it STIS} data,
a new 36 level model of Si has been constructed. This model focuses on
Si~{\sc ii} -- {\sc iv} and contains the ground states of Si~{\sc i} and
Si~{\sc v}, for the sake of the ionisation calculations. The metastable
3p$^{2}$~$^1$D, $^1$S and 3s3p$^3$~$^5$S levels of Si~{\sc i} are also
included. 51 radiative transitions are included in the model (all
treated with Complete Redistribution [CRD] only). The atomic data for
these transitions (gf-values and rates for 
collisional excitation by
free electrons) were extracted from the CHIANTI database. Recombination
rates (radiative plus dielectronic) were taken from Summers (1974a,b, see
Section 2) and photoionisation rates were obtained from the on-line
Opacity Package database (Cunto et al. 1993). Collisional ionisation
rates and charge exchange rates for collisions with both hydrogen and 
helium were taken from Arnaud \& Rothenflug (1985).

\subsubsection{Carbon}

A new 33 level model for C has been constructed to complement the new
Si model. Like the Si model it is focused on the second to fourth
ionisation stages, containing only the ground states of the first and
fifth stages for the ionisation calculations. 52 radiative transitions
are included in the model (all treated with CRD). The atomic data were drawn
from the same sources as used for the Si model.

\subsection{Model A}

Using the assumptions 
mentioned in Section 5.1, DEMD-A was used to compute a new atmospheric model
for the chromosphere/lower transition region.
This model was used by Sim \& Jordan (2003a); the most important atmospheric
parameters ($T_{e}$, $P_{e}$, $P_{H}$) are shown in their fig. 3.

Together with the atomic models discussed in Section 5.2, Model A was used
to compute fluxes for the observed lines of H~{\sc i}, C~{\sc ii -- iv},
Mg~{\sc ii} and Si~{\sc ii -- iv}. These fluxes are given in Table 8, where
 they are compared with the observed fluxes. For comparison, the run of 
electron
temperature against mass column density from the Thatcher et al. (1991)
model was also used to compute fluxes -- these are
given in Table 9. Since the 
Thatcher et al. (1991) model was developed 
using chromospheric lines, only lines of relatively low temperature
ions are listed in Table 9 -- for all of the higher-temperature lines (which 
were outside the scope of their study) the
Thatcher et al. (1991) model substantially overpredicts the fluxes.

In general, model~A can reproduce the observed fluxes of 
Mg~{\sc ii}, C~{\sc iii -- iv} and Si~{\sc iii -- iv} within an accuracy of 
several tens per cent (see Table 8). The poorest agreement for lines of 
these ions is for C~{\sc iii}
977~\AA, the observed flux for which may be subject to significant
uncertainty due to the correction for channel drift in the {\it FUSE}
spectrum.

The greatest failings of model~A relate to C~{\sc ii} and Si~{\sc ii};
although the weak C~{\sc ii} 1010-\AA~multiplet is predicted to within
observational errors (almost), the strong resonance multiplets (1335~\AA~
and 1037~\AA) are overpredicted by factors of $\sim 2$. 
These multiplets are predicted to be significantly
stronger by the radiative transfer calculations (Table 8) than 
the optically thin approximation (Table 7) -- this is due to scattering in the
lines (see Section 5.3.3).
Although 
interstellar absorption affects C~{\sc ii} 1334.5~\AA~ and
1036.3~\AA, it cannot be responsible for the discrepancies in the stronger 
1335.7-\AA~and 1037.0-\AA~lines. In contrast, the strong Si~{\sc ii} 
lines are {\it all} underpredicted by as much as 70 per cent. 
Although
the Thatcher et al. (1991) model does slightly better than model~A for some 
of the chromospheric Si~{\sc ii} lines, it does much less well for 
the hotter transition region lines (i.e. C~{\sc ii}) -- as noted
above, the Thatcher et al. (1991) model was not developed to address 
the transition region in detail.

While small alterations to the DEMD could improve the agreement amongst the
higher temperature lines, it is not possible to simply attribute these 
results to failings in the determination of the DEMD.
By examining the 
emission measure loci in Figs. 2 and 5, it is clear that the 
C~{\sc ii} 
lines form over a wide range $4.0 < \log T_{e}(\mbox{K}) < 4.5$.
To reduce the predicted C~{\sc ii} resonance line fluxes would require 
the DEMD to be reduced in this temperature range. However, this cannot be done
without reducing the predicted fluxes of Si~{\sc iii} or
Si~{\sc ii} which form in the same temperature range.
Such a reduction is unacceptable since the Si~{\sc iii} fluxes are not
overpredicted by model A, and the Si~{\sc ii} fluxes are already underpredicted
to a significant extent.
That simple modifications to the DEMD cannot account for the 
C~{\sc ii}/Si~{\sc ii} discrepancy was checked by constructing $\sim 20$ 
different models with modified DEMDs, none of which were able to 
significantly improve the predicted C~{\sc ii} fluxes without substantial
adverse effects on the Si~{\sc ii} and/or {\sc iii} fluxes.

Therefore, an alternative explanation is required for the 
C~{\sc ii}/Si~{\sc ii} discrepancy. Potential
explanations are discussed below and used in Section 5.4 to construct 
two more models for the outer atmosphere of $\epsilon$~Eri (models B and C).

\begin{table}
\caption{Fluxes computed using MULTI ($F_{c}$) and the 
Thatcher et al. (1991) model compared with observed fluxes
($F_{o}$) for lines of H, Mg, Si~{\sc ii} and C~{\sc ii}. 
$F_{c}$ is given in units of
$10^{-13}$~ergs~cm$^{-2}$~s$^{-1}$. The upper section gives the 
lines from the {\it STIS} spectrum and the lower gives the
those from the {\it FUSE} spectrum.}
\begin{tabular}{lccc}\\ \hline
Ion & $\lambda_{0}$ & {$F_{c}$}&
{$(F_{c} - F_{o})/F_{o} \times 100\%$} \\ 
& (\AA) & & \\ \hline
H~{\sc i}$^{a}$ & 1215.671 & 1175 &639 \\
C~{\sc ii}$^{a}$ & 1334.532& 67.9 &3512  \\
C~{\sc ii} & 1335.68 bl.& 213.9& 4319 \\
Mg~{\sc ii}$^{a}$&2796.352& 398 & 30.5\\
Mg~{\sc ii}$^{a}$&2803.531& 256 & 10.3\\
Si~{\sc ii}&1190.416& 0.377 & 277 \\
Si~{\sc ii}&1193.290& 0.381 & 323\\
Si~{\sc ii}&1194.500& 0.829 & 361\\
Si~{\sc ii}&1197.394& 0.256 & 156\\
Si~{\sc ii}$^{a}$&1260.422& 1.02 & 434\\
Si~{\sc ii}&1264.738&  2.03& 302\\
Si~{\sc ii}&1265.002& 0.470 & 173\\
Si~{\sc ii}&1304.370& 0.303 & 128\\
Si~{\sc ii}&1309.276& 0.455 & 105\\
Si~{\sc ii}&1526.707& 1.07 & 111\\
Si~{\sc ii}&1533.431& 1.53 & 152\\
Si~{\sc ii}&1808.013& 6.26 & -2.0\\
Si~{\sc ii}&1816.928& 11.75 & 12.8\\
Si~{\sc ii}&1817.451& 2.89  & -9.7\\
\\
H~{\sc i}$^{a}$ & 972.537  & 9.70 & 1440\\
H~{\sc i}$^{a}$ & 1025.722 & 29.0 & 692\\
C~{\sc ii} & 1009.858& 1.32 &10900\\
C~{\sc ii} & 1010.083& 2.25 &8233\\
C~{\sc ii} & 1010.371& 2.98 & 6378\\
C~{\sc ii}$^{a}$ & 1036.337& 19.4&16067\\
C~{\sc ii} & 1037.018& 22.1&7495\\ \hline
\end{tabular}

\noindent{$^{a}$ Line affected by interstellar absorption.}

\end{table}

\subsubsection{Atomic data for Si~{\sc ii}}

The limitations of the atomic data used in the radiative transfer calculations
are almost certainly partially responsible for the apparent discrepancy
amongst lines of Si~{\sc ii}. 

The most important uncertainties are in the
electron collision rates. The rates used in the Si~{\sc ii} model are from
Dufton \& Kingston (1991). Although it is difficult to asses the accuracy of
collision rates, Dufton \& Kingston (1991) suggest that their rates should be
accurate to at least $\pm 20$ per cent. Their calculations give rates
between the ground $^2$P levels and all the excited states used in the
Si~{\sc ii} model, and between the metastable $^4$P and excited 3s3p$^2$~$^2$D
levels (which give rise to the 1810-\AA~multiplet), but no collision rates
are provided from the $^4$P levels to other excited states. The absence of
excitation rates from the metastable $^4$P states may mean that the populations
of all the excited states above 3s3p$^2$~$^2$D are underestimated --
this would cause 
the 1530, 1260, 1309 and 1190-\AA~multiplets to be 
systematically underpredicted. Assuming that the
collision strengths from the $^4$P states to levels higher than 
3s3p$^2$~$^2$D are the same as those to the 3s3p$^2$~$^2$D$_{3/2}$ level,
it is found that the predicted fluxes in the multiplets mentioned above are
increased by 53, 25, 73 and 32 per cent, respectively. 
Although only a crude estimate, this suggests that excitation from $^4$P 
could be important and may explain why many of
the Si~{\sc ii} transitions are significantly underestimated by the models.

In addition to direct excitation from the ground state, the 4s~$^2$S level
population
is significantly affected by collisional excitation to 4p~$^2$P followed by 
radiative decay
(Jordan 1969b). 
Thus the computed 1530-\AA~flux is
probably subject to the largest uncertainties due to the atomic data.
The 1810-\AA~flux is subject to the smallest uncertainties, being
dominated by the errors in the collision rates between the low-lying 
levels (at most a few tens per cent). (See also Jordan et al. 2001a.)

\subsubsection{Atomic data for C~{\sc ii}}

The collision rates for C~{\sc ii} are from Blum \& Pradhan (1992) who 
give rates between all levels that are relevant to the formation of the
1335 and 1037-\AA~multiplets. The authors suggested that these rates are
accurate to between 20 per cent and a factor of 2, with greater
uncertainty for higher levels. This suggests that the data should be better 
for the 1335-\AA~multiplet than for the 1037-\AA~multiplet.

\subsubsection{Geometric complications}

The C~{\sc ii} resonance line fluxes are computed to be significantly larger
when radiative transfer is taken into account, compared to the optically thin
limit (compare Tables 7 and 8). This is because the calculations in Section 
4.5 and Table 7 assume that the fraction of photons emitted
in a particular line which escape the atmosphere is $\simeq 1/2$, as
given by the usual 
geometric dilution factor (the optically thin limit). 
For scattering dominated optically thick lines (such as the C~{\sc ii}
resonance lines) that assumption {\it underestimates} the surface flux
since scattering turns around many of the photons that are initially emitted 
downwards into the atmosphere so that they eventually emerge through the upper 
boundary. For a plane-parallel layer this backscattering effect can increase
the flux by up to a factor of two (for a comprehensive discussion of this and 
related effects in the solar case, see Pietarila \& Judge 2004).
In the {MULTI} calculations presented here, the intensity is 
computed at the surface of the star as a function of both wavelength and 
angle accounting for radiative transfer effects (including backscattering)
and this is integrated over angles to compute the flux. Thus 
for an homogeneous, plane-parallel atmosphere, 
the results
presented in Section 5.2 and Table 8 
are more reliable than those
given in Section 4.5. 

In a real stellar atmosphere, however, the situation is more complex; assuming
that $\epsilon$~Eri is like the Sun, the transition region emission will
mostly originate in the bright supergranulation cell boundaries which 
occupy only a fraction of the stellar surface (see, for example the solar
C~{\sc ii} lines discussed and modelled by Judge, Carlsson \& Stein 2003).
We have derived
a filling factor of emitting material in the upper transition region of
$\epsilon$~Eri of $\sim 0.2$ (Sim \& Jordan 2003b). In such a geometry,
much of the emission will not travel solely within the cell boundary material
but will escape sideways into the cell interiors
which are likely to be hotter and less dense. This will alter the probability
that the emission is subsequently scattered upward to escape from the star.
Thus a single-component model may be expected to predict erroneously 
large fluxes 
for lines in which there is significant scattering opacity. (See also 
Jordan, Smith \& Houdebine 2005.) 

The simplest test for this interpretation of the overprediction of the 
C~{\sc ii} resonance line fluxes is to examine C~{\sc ii} lines that are not
affected by back-scattering. Unfortunately the ideal lines for this
examination (the optically thin C~{\sc ii} 2s$^2$2p~$^2$P -- 2s2p$^2$~$^4$P
 lines)
are too weak to measure in the STIS spectrum and all the model predictions
are consistent with this non-detection. The only lines that can be used
for this test are the weak C~{\sc ii} 1010-\AA~lines in the {\it FUSE}
spectrum; the strongest of these lines is measured to an accuracy of
15 per cent. Since they originate in a relatively high-lying energy level, the
atomic data (collision rates) for these lines are likely to be significantly
poorer than for the other C~{\sc ii} lines discussed here. Nevertheless,
these lines are {\it not} significantly overpredicted by model~A (the strongest
line agrees with the observations to within 20 per cent). Since the lines of
the 1010-\AA~multiplet are not affected by back-scattering this supports
the argument that the discrepancy between the observed and computed C~{\sc ii}
resonance line fluxes originates in the treatment of scattering in such lines.

In order to test the hypothesis that the overprediction of the C~{\sc ii}
resonance lines is due to the simplistic 1-D geometry, multi-component
models need to be constructed and used to perform multi-dimensional
radiative transfer calculations. Such work lies beyond the scope of this 
paper. In Section 6 a complete atmospheric model (model~B) is
constructed under the hypothesis that the C~{\sc ii}/Si~{\sc ii} 
discrepancy is
due to errors in the computed C~{\sc ii} resonance line fluxes. 
This model has been used to perform a test (based on the observed Cl~{\sc i}
line) which probes the C~{\sc ii} resonance line strength {\it within}
the atmosphere. That test supports the evidence from the 1010-\AA~multiplet
that the discrepancy lies in the computation of the C~{\sc ii} resonance line
surface fluxes.

\subsubsection{Abundances}

An alternative explanation for the C~{\sc ii}/Si~{\sc ii} 
discrepancy is an error
in the assumed abundances. 
As mentioned in Section 4.1, the relative abundances adopted here are based
on the solar values given by Grevesse \& Sauval (1998). Recently,
Asplund et al. (2004) presented a new set of solar abundances which differ 
in the logarithm from those
of Grevesse \& Sauval (1998) by -0.13, -0.14, -0.17 and -0.04, for C/H, 
N/H, O/H and Si/H, respectively. 
Given that the relative changes of C, N and O are very similar (factors of
 1.35 to 1.48) they will have little influence on the
shape of the EMD deduced based on lines of these elements -- if these new 
abundances were adopted, 
a small global increase in the EMD (equivalent to a small
reduction in the adopted metallicity of $\epsilon$~Eri) would lead to
 computed fluxes which differ only very slightly from those obtained using
 the earlier abundances. Adopting the new abundances, however, would alter
 the relative abundance of C to Si by a factor of 1.23. Although this 
factor would act in the correct sense to address the C~{\sc ii}/Si~{\sc ii} 
problem here, it is insufficient to explain most of the discrepancy. 
Therefore, the standard Grevesse \& Sauval (1998) abundances will be retained
 throughout this investigation (although the influence of adopting the 
Asplund et al. 2004 abundances will be quantified in Section 6.3).

To bring the
C~{\sc ii} resonance lines and Si lines into agreement, either the C abundance
could be made smaller or the Si abundance made larger. If the assumed
abundances were wrong, one would expect to see a discrepancy between the
lines of higher ions (C~{\sc iii}, {\sc iv} or Si~{\sc iv}) and lines of other
species. Unfortunately, neither C~{\sc iv} nor Si~{\sc iv} are ideal for
comparison since they are very sensitive to the ionisation balance 
calculation (Section 3).
The observed O~{\sc iii} lines form at similar temperatures to the C~{\sc iii}
1175-\AA~multiplet and the Si~{\sc iv} lines. Based on the multiplet fluxes in
Table 7, there is no obvious discrepancy between O~{\sc iii} and C~{\sc iii}.
Si~{\sc iv} is different, however, being underpredicted rather than 
overpredicted; thus it is concluded that if abundance variations are
responsible for the C~{\sc ii}/Si~{\sc ii} it is 
the Si abundance that is likely to be in error.

In order to remove the C~{\sc ii}/Si~{\sc ii} 
discrepancy, the Si abundance needs to be
increased by about a factor of 2.4, relative to the value adopted in
model~A. 
The interpretation of such an increase requires some care.
Abia et al. (1988) have measured the photospheric Si abundance in
$\epsilon$~Eri. They found the logarithmic Si abundance relative to
the Sun to be $[ \mbox{Si} /\mbox{H} ] = -0.11$~dex. 
Based on their discussion of
likely errors, the uncertainty in this value should not be more than
$\pm 0.25$~dex. 
More recent studies have found similar results: 
$[ \mbox{Si} /\mbox{H} ] = -0.10$ and $-0.04$ 
(Bodaghee et al. 2003; Zhao et al. 2002, respectively)
with much smaller errors
($\sim 0.05$~dex).
These measurements are all consistent with the assumed
metal abundances in model~A which are based on Drake \& Smith (1993).
It suggests, therefore, that a proposed increase in the transition 
region Si abundance would be an increase {\it relative to the stellar
photospheric abundance}. Note that Zhao et al. (2002) deduce a
C abundance which is {\it slightly higher} than solar and thus, if adopted,
would make the discrepancy worse. However, since their C analysis
involves only one spectral line the error in this result may be significantly
larger than that in the Si abundance.

It is concluded that if abundance variations are responsible for the
C~{\sc ii}/Si~{\sc ii} discrepancy, the Si abundance enhancement
required is relative to the $\epsilon$~Eri photospheric
abundance.
There is considerable interest in the relative abundance of elements 
in the transition regions and coronae of cool stars. Many authors
(see for example Jordan et al. 1998 and references therein), have discussed
the possibility that the relative abundances of elements in the corona are
different from those in the photosphere to a degree which correlates
with the {\it First Ionisation Potential} (FIP) of the element.
It is suggested that, in the upper atmosphere, 
elements with high FIPs ($> 10$~eV) have lower
abundances relative to those with low FIPs. 
The first evidence of this ``FIP-effect'' was found in the solar wind 
(Crawford, Price \& Sullivan 1972).
The physical origin of the effect is unknown.
Based on studies of the
solar atmosphere, Feldman et al. (1992) give coronal elemental abundances,
assuming that the FIP-effect acts to increase the low-FIP element
abundances. They give a relative coronal abundance for Si to C which is a 
factor 3.8 greater than the photospheric relative abundance.
Laming, Drake \& Widing (1995) used full disk solar
observations to confirm the coronal abundances given by Feldman et al. (1992).
Their work found evidence of the FIP-effect in the corona, but they found
no suggestion of a FIP-effect operating at lower temperatures
($\log T_{e}(\mbox{K}) < 6$). 
Jordan, Macpherson \& Smith (2001) 
studied spatial variations in the Si~{\sc iii}/C~{\sc ii}
intensity ratio across regions of the quiet Sun and a coronal hole.
They found no differences in excess of 0.2~dex.
Since the FIP effect is not thought to operate in coronal holes this 
suggests that the FIP enhancement of Si in the quiet Sun is less than a 
factor of 1.6.
These results are surprising. Since the first ionisation for most
elements takes place at chromospheric temperatures,
one would expect an effect 
associated with the first ionisation potential to originate in the 
chromosphere and thus to be apparent in the transition region,
but the solar work suggests that  
the effect only occurs at significantly
higher temperatures.

Several studies have examined the case for a FIP-effect in other cool stars.
Laming, Drake \& Widing (1996) looked for a FIP-effect in the
corona of $\epsilon$~Eri. They found some evidence for a solar-like
FIP-effect, but were unable to make a definitive detection.
More recently, Drake, Laming \& Widing (1997) 
and Laming \& Drake (1999) have identified
solar-like FIP-effects in the coronae of $\alpha$~Cen and $\xi$~Boo.
It has also been found, however, that there is no apparent FIP-effect in the
corona of Procyon (Drake, Laming \& Widing 1995). 

In summary, to date, there is no compelling evidence for the operation of
a FIP-effect in the chromospheres/transition regions of the Sun or
any other late-type stars. If there is such an effect 
in the coronae of stars,
however, it is difficult to understand why it would not occur in the 
lower parts of the atmosphere.
The C~{\sc ii}/Si~{\sc ii} 
discrepancy in the modelling of $\epsilon$~Eri could
be interpreted as evidence for a FIP enhancement of the Si abundance in
the low atmosphere of up to a factor of 2.4, 
and an atmospheric model based on that assumption is
constructed in Section 6 (model~C). There are, however,
significant uncertainties and several shortcomings for this explanation
which are also discussed in Section 6.

\section{Models B and C}

In this section, the discussion of model~A (Section 5.3) is used to develop
two more models (B and C). These models were constructed in the same manner
as model~A (see Section 5.1) but are based on different EMDs for the 
transition region and include a small amount of macroturbulence.
The differences between models B and C
relate to the possible interpretations of the C~{\sc ii}/Si~{\sc ii} 
discrepancy
identified in model~A (see below).
The complete atmospheric models (A, B and C) are tabulated in 
Appendix~B.

\subsection{Modifications to the turbulent velocity field}

For model~A it was assumed that all the observed non-thermal motion was due to
microturbulence. 
Since the optically thick chromospheric line profiles are
sensitive to the distinction between macroturbulence and microturbulence,
the radiative
transfer calculations for model~A can be used to improve the prescription for
the turbulence in the new models~B and C.

\begin{figure*}
\epsfig{file=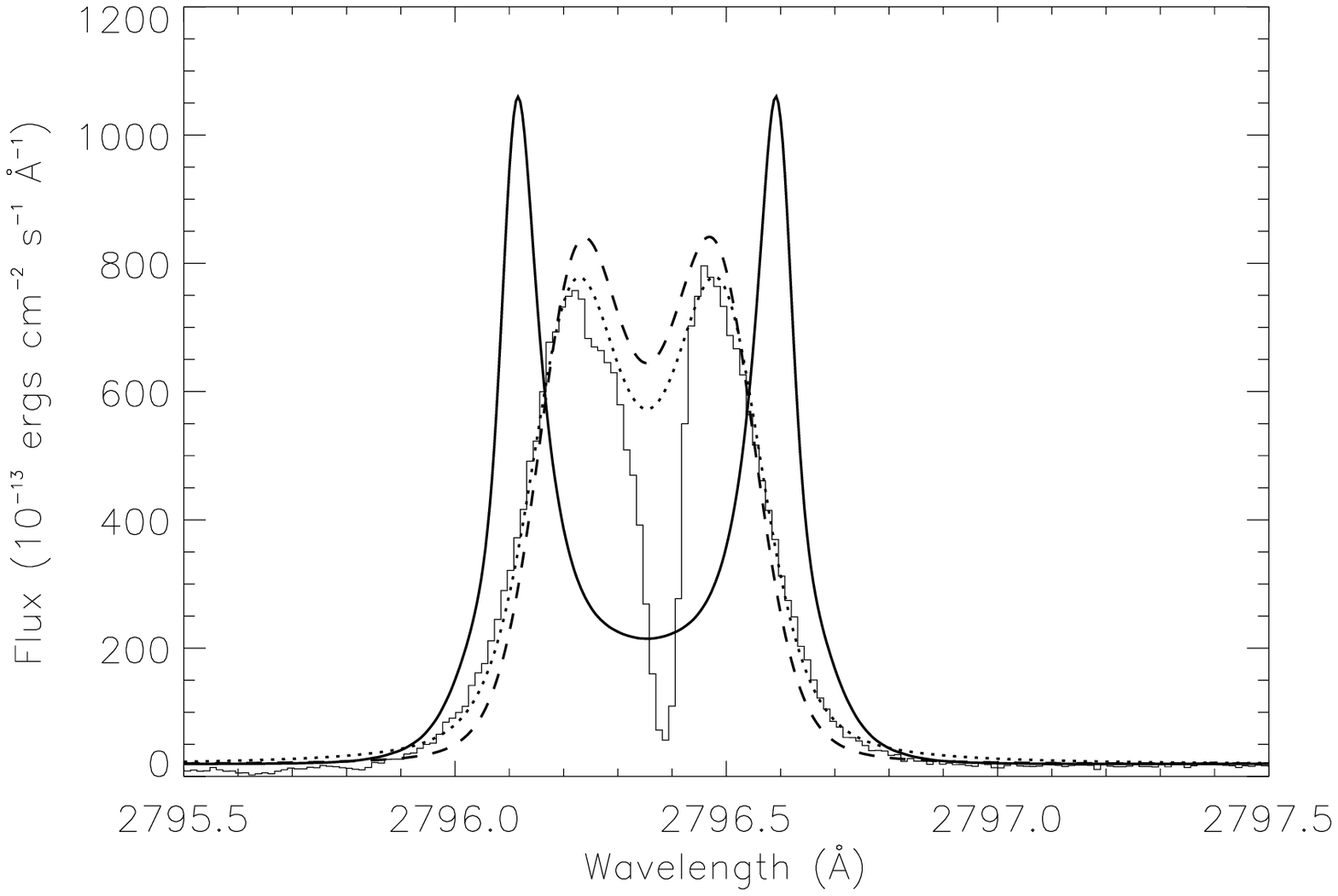, width=6cm}
\epsfig{file=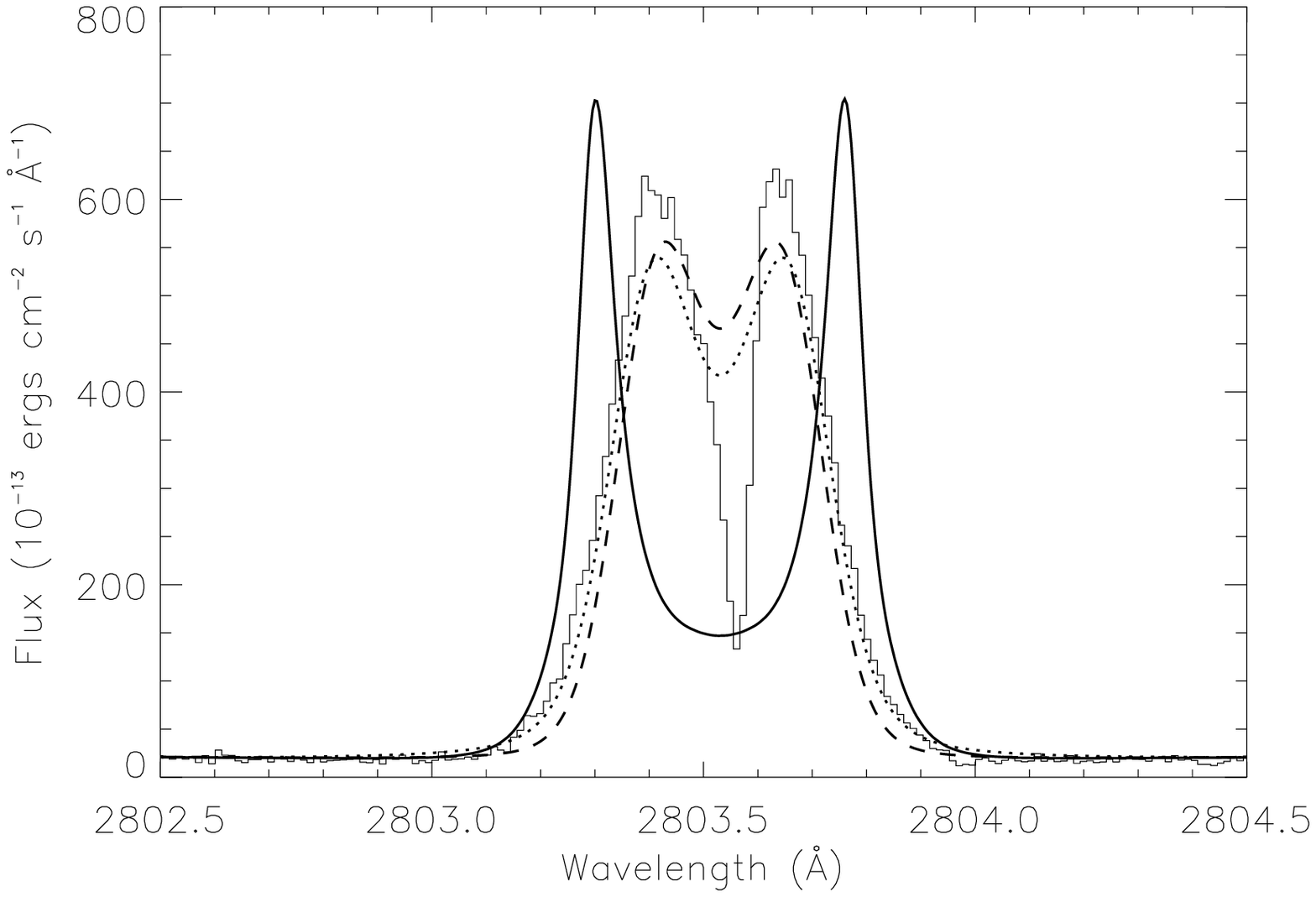, width=6cm}
\caption{Comparison of computed and observed Mg~{\sc ii} profiles. In each
figure the solid histogram shows the data, the solid curve shows the 
model~A calculations, the dashed curve shows the model~B calculations and
the dotted curve shows the model~C calculations.
The narrow dips in the observed profiles are due to interstellar absorption.}
\end{figure*}

In Fig.~6, the computed Mg~{\sc ii} h and k lines are
plotted with the observed line profiles from STIS. Clearly the central 
reversal in the model~A profiles is too wide, indicating that the assumed
microturbulence is too large. To compensate for this, macroturbulence is
invoked to explain some of the observed line width: by trial and error, it
was found that a value of $\xi_{\mbox{\scriptsize macro}} = 8.5$~km~s$^{-1}$
leads to much better profiles for the Mg~{\sc ii} lines. Unfortunately, 
there are no suitable lines to derive the macroturbulence in other parts of 
the atmosphere and so it is simply assumed that the macroturbulence does
not vary through the upper chromosphere and transition region: thus for
models~B and C, the microturbulent velocity field used above $\log T_{e}[K]
= 3.8$ is

\begin{equation}
\xi^{2}_{\mbox{\scriptsize micro}}(T_{e}) = \xi^2 (T_{e}) - \xi^{2}_{\mbox{\scriptsize macro}}
\end{equation}
where $\xi(T_{e})$ is the non-thermal velocity specified by equation~(3).
The adopted value for the
macroturbulence has little significance above the low transition region, where 
it leads to less than 10 per cent difference in the adopted microturbulence.

\subsection{Modifications to the EMD}

For both models~B and C the EMD has been slightly modified at high temperatures
to improve the agreement between some of the predicted and observed
line fluxes. In addition, more significant modifications have been made at 
low temperatures; these modifications are discussed further below. The
complete EMDs for the models are compared with that for model A in Fig. 7, and the 
atmospheric models obtained from the EMDs are compared with model~A and the earlier models
by Thatcher et al. (1991) and Kelch (1978) in Fig. 8.

\begin{figure*}
\begin{center}
\epsfig{file=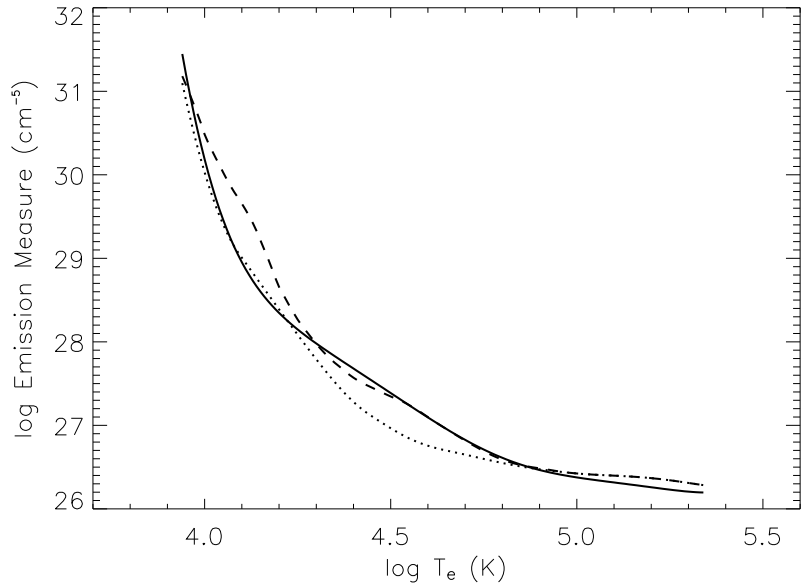, width=12cm}
\end{center}
\caption{Comparison of mean emission measure distributions. The solid
curve is for model~A, the dashed curve
is for model~B  and the dotted curve is for model~C. 
In each case the mean emission
measure $Em^{0.3}(T_{e})$ is plotted.}
\end{figure*}

\begin{figure*}
\begin{center}
\epsfig{file=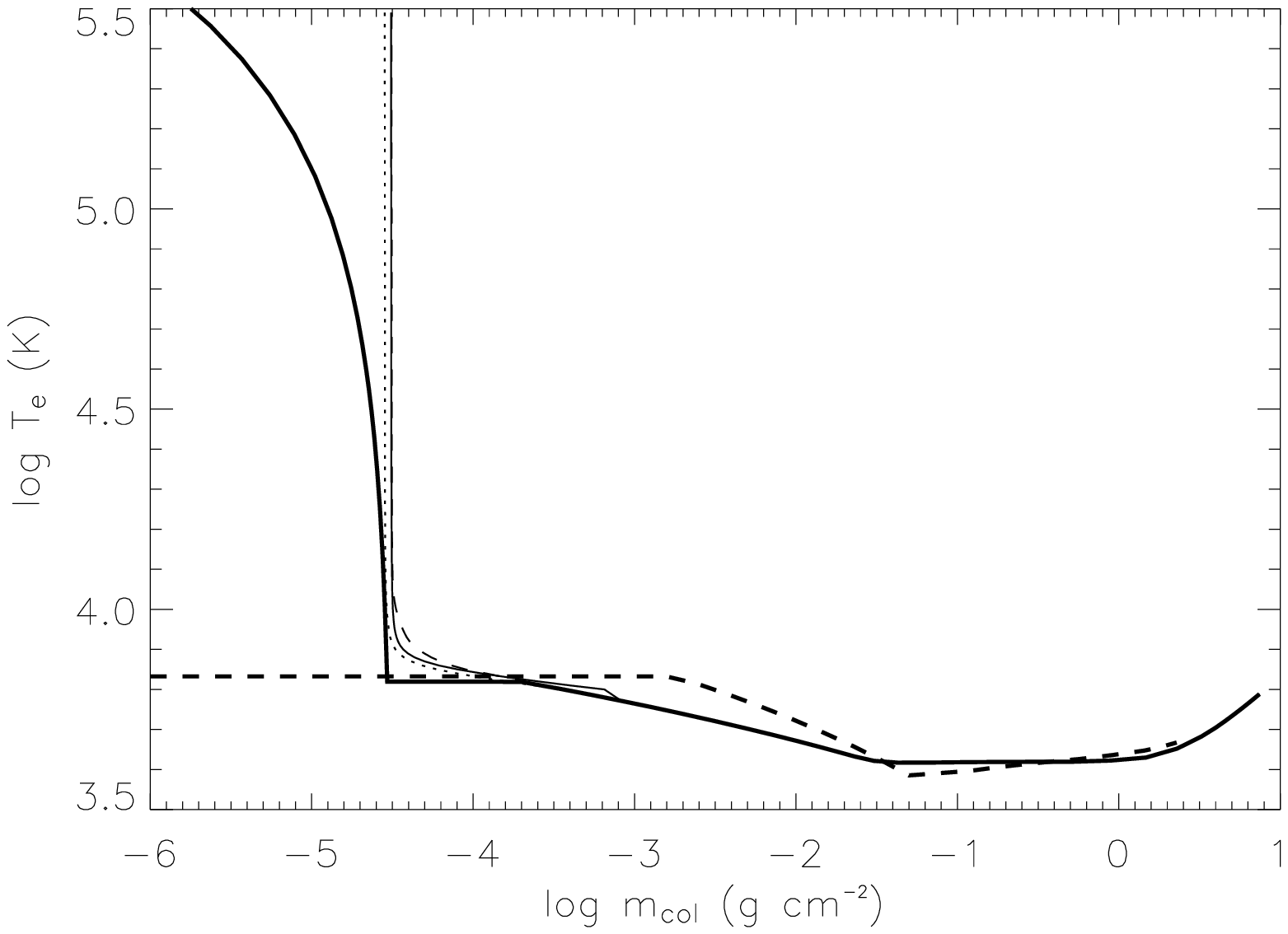, width=6cm}
\epsfig{file=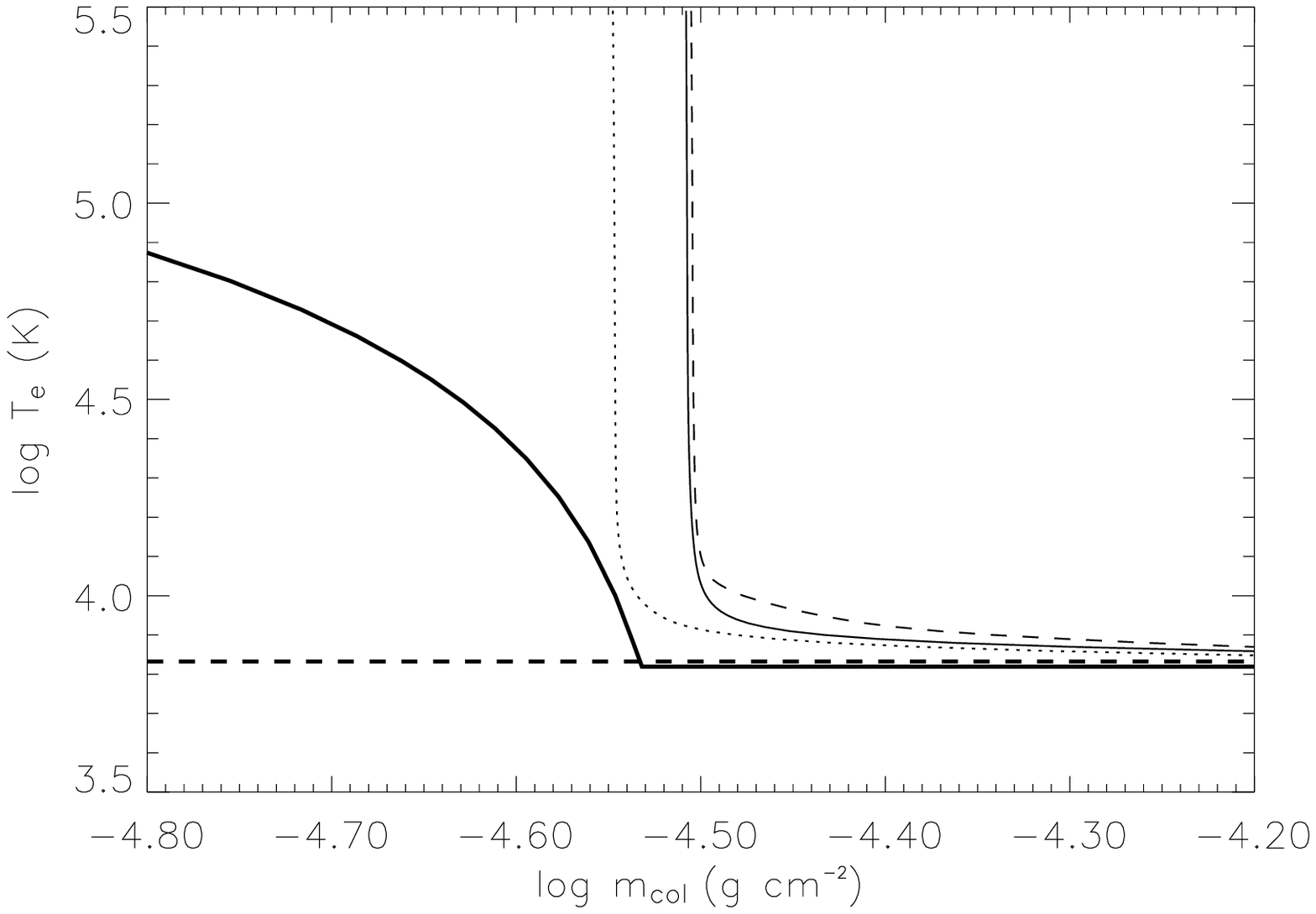, width=6cm}
\end{center}
\caption{Comparison of atmospheric models (column mass density versus electron temperature).
The left panel shows the complete models and the right panel shows the transition region in
detail. The heavy dashed line is the Kelch (1978) model, the heavy solid line is the
Thatcher et al. (1991) model, the light solid line is model~A, the light dashed line is model~B and
the dotted line is model~C.}
\end{figure*}

\subsubsection{Model B}

Model B has been constructed based on the assumption that the discrepancy 
between the computed and observed C and Si fluxes is the result of errors in 
the computed C~{\sc ii} resonance line surface fluxes, primarily due to the
simplistic one-dimensional geometry used here (see Section 5.3.3). For this model, the
DEMD was significantly increased around $\log T_{e}[K] = 4.0$ and slightly 
reduced
at higher temperatures. These alterations increase the 
computed Si~{\sc ii} fluxes and slightly reduce the C~{\sc ii} and
Si~{\sc iii} fluxes relative to model~A.

\subsubsection{Model C}

Model C has been developed to investigate the alternative explanation of
the C~{\sc ii}/Si~{\sc ii} 
discrepancy in model~A, that of a (possibly FIP-related) 
change in the Si abundance. For this model, the assumed Si abundance was
increased by a factor of 2.4. 
If this increase is
due to a FIP-related effect then one would expect that the abundances of
other low-FIP elements should be increased by a similar factor. Of the
other element considered here this affects Mg  and Al -- hence the Mg and Al
abundances have also been raised by a factor of 
2.4 relative to those in model~A.

The DEMD for model C was reduced throughout
the region $\log T_{e}[K] < 4.5$ (compared with DEMD-A) 
to allow for the change in the emission
measure loci resulting from the higher low-FIP-element abundances.

\subsection{H Lyman line profiles}

Since the cores of the strong H~{\sc i} Lyman lines are heavily affected 
by interstellar absorption, the
diagnostic power of these lines for studying the stellar atmosphere lies 
in examining the line wings.
Fig.~9 compares the Lyman~$\alpha$, $\beta$ and $\gamma$ line profiles 
computed from models B and C with
the observations. The model profiles have been corrected for interstellar 
absorption using the H~{\sc i} column
densities and velocities derived for the $\epsilon$~Eri line-of-sight by Dring
 et al. (1997).

For Lyman~$\beta$ and $\gamma$, models B and C both predict the profiles
 very well. Model B tends to slightly
overpredict the line strength but only by 10 -- 20 per cent. For 
Lyman~$\alpha$ the agreement is less good -- both
models predict profiles with significantly weaker line wings than observed. 
This suggests that the models 
underestimate the importance of photon scattering from the line core to the
 wings. Since this effect is not
evident in the Lyman~$\beta$ or $\gamma$ profiles (nor the Mg~{\sc ii} 
profiles) it seems unlikely that this is
solely the result of errors in the atmospheric velocity field -- most 
likely it indicates the coherence
fraction used in the PRD treatment of this line is too high. A detailed
investigation of the Lyman~$\alpha$ profile goes beyond the scope of this
 work, but
to illustrate the influence of a smaller coherence fraction, Fig.~9
also shows the Lyman~$\alpha$ profile computed from model~B adopting a 
frequency-independent, depth-independent 
coherence fraction of $\gamma = 0.9$ (this is in contrast to the other 
calculations which use the physical 
$\gamma$ implemented in the Hubeny \& Lites 1995 code). As expected, the 
constant-$\gamma$ calculation predicts stronger line wings bringing it into 
much better agreement with the data. However, there are still some
 significant differences in the line shape which
suggest that more redistribution in needed in the line core but less to 
the far wings. 
The illustrative $\gamma$ adopted is rather low compared to theoretical 
calculations and lower than the preferred
Lyman $\alpha$ maximum coherence-fraction suggested by Vernazza et al. 
(1981) for their solar models
($\gamma = 0.98$). However, when comparing with the Vernazza et al. (1981)
 value, it is noted that $\gamma$ is expected to be lower in $\epsilon$~Eri
than the Sun since the density is higher.

\begin{figure*}
\epsfig{file=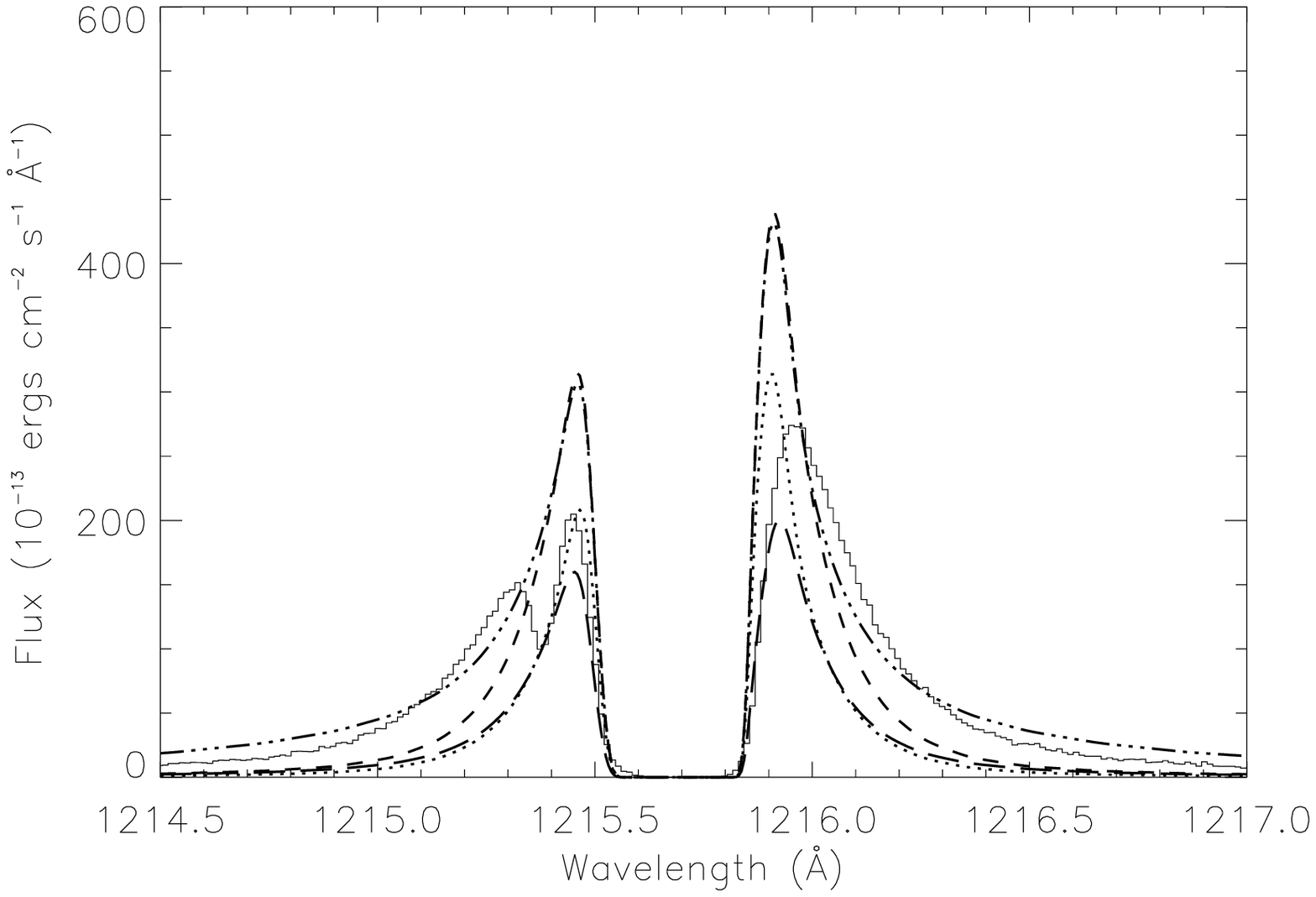, width=5cm}
\epsfig{file=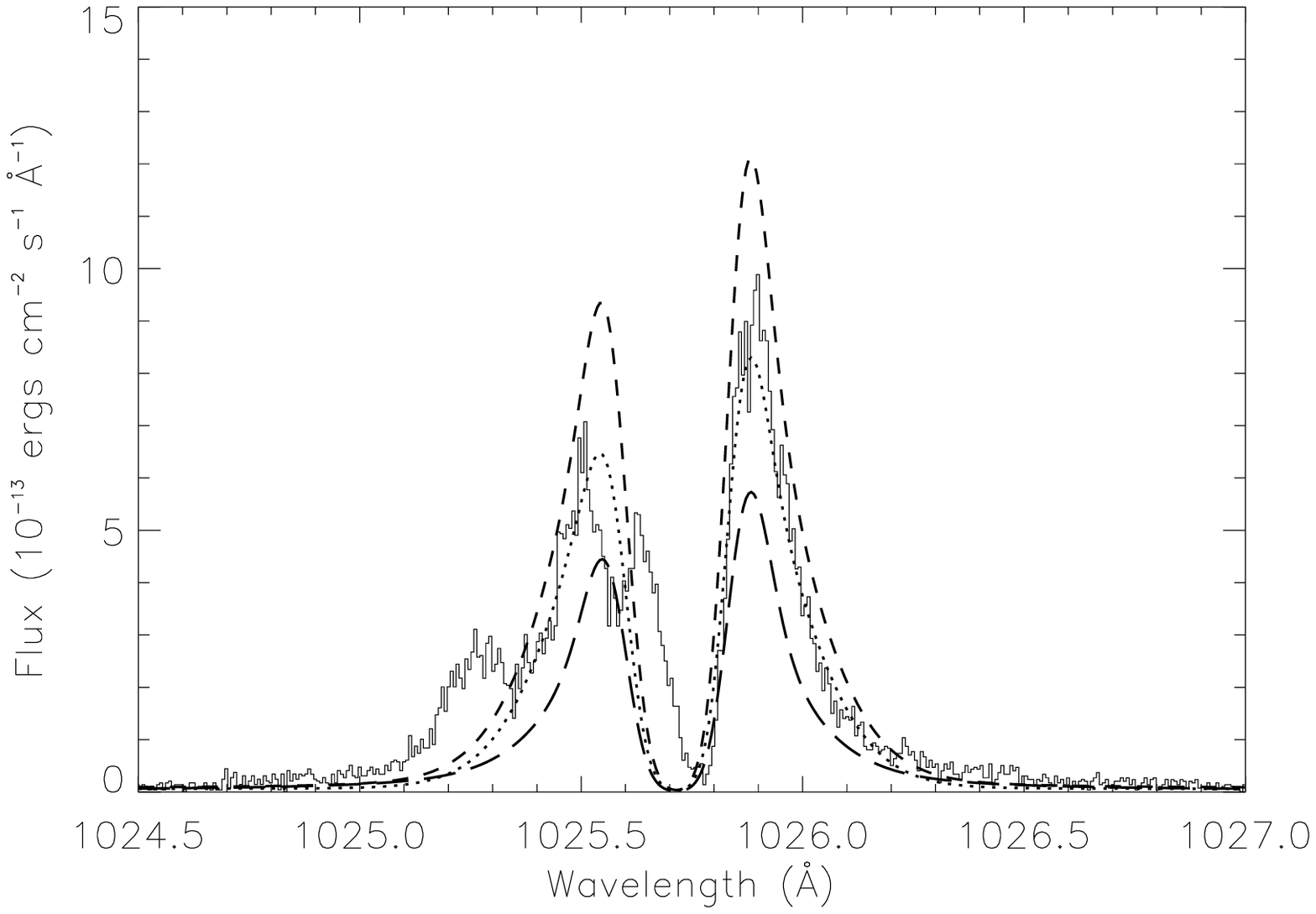, width=5cm}
\epsfig{file=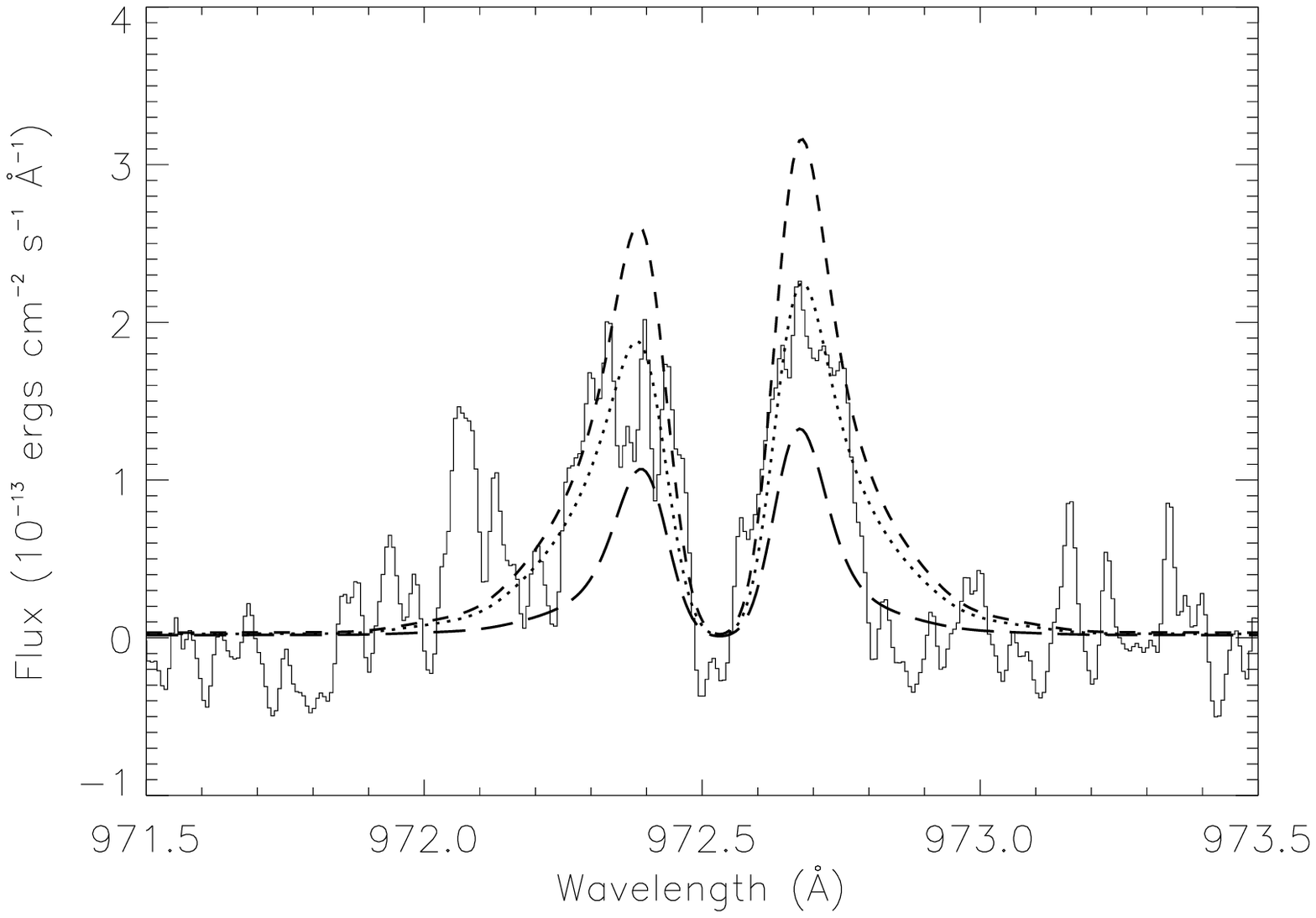, width=5cm}
\caption{Comparison of computed and observed H~{\sc i} profiles. In each
figure the solid histogram shows the data, the dashed curve shows the model~B calculations,
the dotted curve shows the model~C calculations and the long-dashed curve shows the model~D 
calculations.
In the left panel (Ly~$\alpha$), the triple-dot-dash curve shows the model~B result if a 
frequency-independent, depth-independent coherence fraction of 0.9 is adopted (see text).
}
\end{figure*}

\subsection{Computed fluxes}

The new models (B and C) have been used to compute fluxes for all the lines
considered for model A. Table 8 gives the fluxes computed from radiative
transfer calculations and Table 10 gives the fluxes for the high temperature 
multiplets for which radiative transfer calculations have not been performed
(the model fluxes for these lines are obtained directly from the DEMD on the
assumption that the multiplets are effectively optically thin). In both 
tables, 
the model B fluxes are followed (in parenthesis) by fluxes computed from a
 model
identical to model B saving that it adopts the Asplund et al. (2004)
 relative abundances rather than those of Grevesse \& Sauval (1998).

\begin{table}
\caption{Comparison of calculated fluxes ($F_{c}$) and observed fluxes ($F_{o}$)
for emission measure distribution models~B and C. Following the model B fluxes,
values are given for computations from a model identical to model B saving that the
Asplund et al. (2004) relative abundances are adopted. 
$F_{c}$ given in units of $10^{-13}$~ergs~cm$^{-2}$~s$^{-1}$.
The top part of the table gives the density-insensitive lines,
the middle gives the density-sensitive line
and the lower part gives the {\it FUSE} lines.}
\begin{tabular}{lccccc} \hline
Ion & $\lambda$ (\AA)& \multicolumn{2}{|c|}{$F_{c}$} & \multicolumn{2}{|c|}{$(F_{c} - F_{o})/F_{o} \times 100\%$} \\ 
& & Model B & Model C & Model B& Model C\\ \hline
N~{\sc v} & 1240 &  1.27 (0.92) &  1.27 &    -10 (-34) & -10 \\
O~{\sc v}$^a$ & 1371 &  0.078 (0.053) & 0.078 &      -25 (-50) & -25 \\
Al~{\sc iii} & 1854 & 0.204 (0.162) &  0.218   &     -60 (-68) & -57\\
S~{\sc ii} &1255&   0.190 (0.123) &   0.100    &    -2 (-37)&-48\\
S~{\sc iv} & 1406&  0.090 (0.058) &   0.088     &    21 (-21) &19\\ 
\\
N~{\sc iv} & 1486&  0.069 (0.050) &0.068&         52 (11) &52\\
O~{\sc iii} & 1663& 0.213 (0.144) &  0.196&       19 (-17) &9\\
O~{\sc iv} &1400& 0.815  (0.551) &  0.814&     129 (55) &129\\
O~{\sc v}$^a$ &1218& 0.671 (0.454) & 0.671&       -1 (-33) &-1\\ 
\\
N~{\sc iii} & 991 & 0.452 (0.327)&   0.409&      37 (-1.0) & 24 \\
O~{\sc vi}$^a$& 1035 & 3.94 (2.66) &     3.94&        -43 (-61) & -43\\ 
S~{\sc iv} & 1070 & 0.240 (0.155) &    0.238&        156 (65) & 153 \\
S~{\sc vi} & 933 & 0.100 (0.065)&     0.100&   -58 (-73) & -58 \\ \hline
\end{tabular}

$^a$ Flux likely to be underestimated since the line forms partially outside
the temperature range covered by the models.

\end{table}

The observed Cl~{\sc i} 1351-\AA~line 
provides a useful probe of the C~{\sc ii} 
1335-\AA~radiation field {\it within} the atmosphere since the Cl~{\sc i} line
forms by fluorescence, pumped by the 1335-\AA~transition (Shine 1983).
A very simple three level Cl~{\sc i} model has been used to compute the
1351-\AA~flux. The model contains only the ground configuration 
$^2$P$^{\mbox{\scriptsize o}}$ states
and the excited 3p$^4$($^3$P)4s~$^2$P$_{1/2}$ level. The only physical
processes included in the model are the two radiative transitions
$^2$P$_{3/2}^{\mbox{\scriptsize o}}$ -- $^2$P$_{1/2}$ 
(the transition pumped by C~{\sc ii}) and the
observed transition ($^2$P$_{1/2}^{\mbox{\scriptsize o}}$ -- $^2$P$_{1/2}$).
The oscillator strengths were taken from Kurucz \& Bell (1995).
For the calculation the C~{\sc ii} radiation field was taken from MULTI
calculations with the full C model (Section 5.2.4).
Table 11 gives the computed Cl~{\sc i} flux for both models~B and C. It is
to be expected that the Cl~{\sc i} flux may be overestimated since no
levels of ionised Cl~{\sc i} are included in the calculation (i.e. it is 
assumed that the Cl~{\sc i} ion fraction is 1.0).

\begin{table}
\caption{Comparison of calculated fluxes ($F_{c}$) and observed fluxes ($F_{o}$)
for Cl~{\sc i} in models~B and C. 
$F_{c}$ given in units of $10^{-13}$~ergs~cm$^{-2}$~s$^{-1}$.}
\begin{tabular}{lccccc} \hline
Ion & $\lambda$ (\AA)& \multicolumn{2}{|c|}{$F_{c}$} & \multicolumn{2}{|c|}{$(F_{c} - F_{o})/F_{o} \times 100\%$} \\ 
& & Model B & Model C & Model B& Model C\\ \hline
Cl~{\sc i} & 1351.656 & 0.231 & 0.110 & 16.7 & -80.0 \\ \hline
\end{tabular}
\end{table}

\subsection{Discussion}

The computed fluxes given in Tables 8, 10 and 11 are now used to 
assess the successes, failures and relative merits of models
B and C. 
The following points are of particular note:

\begin{enumerate}

\item In general, both models B and C predict fluxes for the high
temperature lines (Si~{\sc iii} and hotter) that agree well with the 
observations, given the combined uncertainties in the measured fluxes
and atomic data (in particular the ionisation balance).

\item The Mg~{\sc ii} lines are predicted well by both models, both in
terms of line profile and flux. The flux ratio is slightly 
different from that observed, possibly indicating that there is
insufficient opacity in the models. It should be noted that if 
inhomogeneity
affects the C~{\sc ii} lines (as postulated for model B), 
the Mg~{\sc ii} lines may also be affected -- but without 
multi-dimensional calculations the scale and nature of such an effect is
unknown.

\item The Si~{\sc ii} 1810-\AA~multiplet is well predicted (within
20 per cent) by both models. This is the most important of the Si~{\sc ii}
multiplets since it is the least sensitive to the absence of
collision rates between the $^4$P and more highly excited energy levels.

\item The higher excitation Si~{\sc ii} multiplets are predicted much less
accurately. However, as discussed in Section 5.3.1 these may be significantly
affected by inadequacies in the Si~{\sc ii} atomic data which may account
for differences of 50 per cent or more.

\item The C~{\sc ii} resonance line fluxes (1335~\AA~and 1037~\AA)
are significantly overestimated by model B ($\sim $ a factor of two)
but predicted quite well by model C. As discussed in Section 5.3.3 the
overprediction may be the result of the simplistic geometry assumed in 
the models.

\item Although the uncertainties are large, the C~{\sc ii} 1010-\AA~
multiplet fluxes are {\it not} significantly overpredicted by model~B.
They are slightly underpredicted by model~C.

\item The Cl~{\sc i} fluorescent flux is significantly {\it underpredicted}
for model C. Model B predicts this line flux well.

\item The S~{\sc ii} flux is more accurately predicted by model~B than C.

\item If the Asplund et al. (2004) relative abundances are adopted, the model
 C~{\sc ii} resonance
line fluxes are closer to those observed but are still overpredicted. 
Adopting the new abundances does not significantly improve the agreement 
amongst the high-temperature lines since the
relative abundances of C to N and O have not changed very much from those 
given by Grevesse \& Sauval (1998).

\end{enumerate}

As discussed previously, most of the exceptions to (i) can be explained
by inadequacies in either the ionisation balance for alkali-like ions
(Al~{\sc iii}, S~{\sc vi}) or the atomic data. However, the tendency of the
density-sensitive lines to be overpredicted may also indicate the presence 
of density inhomogeneities and, therefore, the need for multi-component 
modelling (see Section 4.3).

Points (vi), (vii) and (viii) all argue in favour of model B over model C.
The only significant advantage of model C over model B is that it predicts
the C~{\sc ii} resonance line surface fluxes more accurately. However, 
model C has one more free parameter than model B (the FIP-enhancement factor)
and so it is not surprising that better agreement can be found. Point (vii)
is particularly crippling for model C since it suggests that the model
does not reproduce the C~{\sc ii} radiation field {\it within} the atmosphere.

Thus it is concluded that model~B is the better single-component model
for the upper transition region of  $\epsilon$~Eri, and that its main 
limitation is the assumption of homogeneity. In future work 
inhomogeneous models and multi-dimensional radiative transfer must be
employed to resolve the issues relating to scattering in the C~{\sc ii}
resonance lines.

The interpretation of the apparent discrepancy in the C~{\sc ii}/Si 
flux ratios in terms of
a variation in the abundance ratio (model C) is judged to be inferior. Model C
is, however, a useful constraint on FIP-related abundance
enhancement in the transition region of $\epsilon$~Eri: 
any enhancement in the Si abundance is constrained to be no larger than that
adopted in model C (a factor of 2.4).

\section{A simple multi-component atmospheric model}

It is well known from direct observations of the Sun and studies of other
stars (e.g. the K giant $\alpha$~Tau, McMurry \& Jordan 2000) that the atmospheres
of cool stars are inhomogeneous.
As mentioned in Section 5.3.3, Sim \& Jordan (2003b) concluded that 
emitting material occupied only a limited fraction ($\sim 20$ per cent in the
mid-transition region) of the available volume in the atmosphere of
$\epsilon$~Eri. 
All the models discussed above
(A, B and C) are single-component and assume a filling factor of 100 per 
cent. 
One important consequence of a small filling factor is that line-of-sight 
opacities will be higher since the plasma is confined by a smaller
cross-sectional area.
To fully investigate multi-component atmospheric models requires 
multi-dimensional radiative transfer which goes beyond the scope of this 
paper, however, in this section a crude two-component model will be 
presented and contrasted with the homogeneous models discussed above.

For the two-component model (model D) it is assumed that the atmosphere has a 
uniform filling factor of emitting material, fixed at 20 per cent (i.e. 
$A(r) / A_*(r) = 0.2$). The apparent EMD of the emitting material is 
the same as that used for model B.
For model D the intrinsic EMD
is therefore larger by a factor of 5 and hence leads to larger line
opacities.
The remainder of the volume is assumed to be filled by a
completely dark second component. One-dimensional radiative transfer 
calculations are performed
for the emitting component of the model neglecting the dark component. 
This model can only be regarded as exploratory since it neglects the
probable process of photon leakage from the emitting component to the dark 
component but it is useful to investigate the effects of higher line 
opacities.

Fluxes computed from 
radiative transfer calculations with model D are given in Table~8. 
Since the EMD for model D is identical to that of model~B, the model~D
fluxes for lines assumed to be optically thin (Table~10) are the same as
those for model B. The Cl~{\sc i} flux for model~D is also virtually 
identical to that of model~B (Table~11).

Comparing the
model B and D fluxes in Table~8 reveals several interesting changes,
mostly in resonance lines that form at relatively low temperatures (these being
the lines for which opacity effects are most important). 
The most dramatic differences are in the absolute
fluxes of the strongest lines, those of H~{\sc i} and Mg~{\sc ii}. Increased
 opacity
reduces the fraction of emitted photons which eventually escape the 
atmosphere in these
lines. The effect is greatest in the H Lyman lines where the model D 
fluxes are reduced by
 a factor $\sim 2$, relative to those from model B (see Fig.~9). Although 
this shows
that the 1-D models presented here (B and C) predict the Lyman lines more
 accurately than the 
two-component model (D), it must be noted that the very simple 
implementation of model~D 
neglects photon escape from the emitting component to the second component.
Therefore it is expected that model~D will underpredict optically thick 
line fluxes relative to
a more sophisticated, multi-dimensional radiative transfer calculation for 
a multi-component atmosphere.

The Mg~{\sc ii} lines are less affected than those of hydrogen but still
 decrease 
significantly in strength (by a factor of $\sim 1.6$). The computed ratio
 of the two Mg~{\sc ii}
lines changes in the sense which brings it closer to the observed ratio, 
but the change is very small.

For lines of all other ions, the changes
in absolute flux are quite small, at most $\sim$ 10 -- 20 per cent.
For some pairs of Si~{\sc ii} lines with common upper levels (i.e. the
1526.7/1533.4-\AA~pair and, in particular, the 1808.0/1817.5-\AA~pair) the
line ratios from model D are in better agreement with the data that those from
model B. The provides some evidence that the line-of-sight opacities
 suggested by
the two-component model are more realistic than those in the homogeneous 
models.
However, the 1309.3/1304.4-\AA~flux ratio from model D agrees less well 
than that from model B (note, however, that the absolute fluxes for these lines
are generally in poorer agreement with the data than those of the other lines 
considered here).

The
C~{\sc ii} 1334.7/1334.5-\AA~flux ratio changes from 2.55 in model B to 
2.71 in model D, both
these values being consistent with the observations, while the 
1036.3/1037.0-\AA~ratio is in 
substantial disagreement with the data for both models. 

It is concluded that the simple two-component model presented here 
is not significantly superior to the homogeneous models discussed above, 
although there are
some improvements for the modelling of the  most important Si~{\sc ii}
 line ratios. A more complete treatment incorporating
multi-dimensional radiative transfer effects is needed to fully investigate
 a multi-component model; this is deferred to later work.

\section{Summary and conclusions}

Using measured ultraviolet emission line fluxes and widths, new 
single-component
models for the upper chromosphere/transition region of
$\epsilon$~Eri have been constructed; a preliminary model (model A)
and two further models (models B and C)
which address some of the inadequacies of the first model. To date, 
these models are the most detailed that have been constructed for the 
chromosphere/transition region 
of a late-type main-sequence star, other than the Sun.
A simple two-component model (model D) has also been presented. The 
predictions of that model were only subtly different from those of the 
equivalent single-component model, suggesting that the 1-D modelling does
provide useful average models.

For the emission measure analysis involved in the construction of the models
new ionisation balance calculations have been performed which
attempt to account for the density sensitivity of dielectronic
recombination rates. Although not 
definitive, these calculations highlight the extreme sensitivity of the 
analysis of certain ions (generally the alkali-like ions) to errors in the
recombination rates and illustrate the need for the publication of reliable
recombination rates at finite density.

In most respects, line fluxes/profiles computed with 
the favoured model (model B) are in agreement with the observations,
allowing for modest measurement uncertainties and 
reasonable errors in the adopted 
atomic data and ionisation/recombination rates. 
The most significant failures of the model are its inability to 
compute reliable fluxes for the optically thick resonance lines of 
C~{\sc ii} and the spin-forbidden transition region lines of several
ions (most notably O~{\sc iv}). Barring large errors in the atomic
data, the most probable
explanation for the discrepant transition region forbidden
line fluxes lies with density variations in the atmosphere. The C~{\sc ii}
resonance lines have been examined in detail and used to place a
constraint on possible FIP-related abundance variations within the upper 
atmosphere.
However, it has been concluded that the anomaly associated with these lines
is most likely to be related to scattering in an 
inhomogeneous atmosphere.

Thus the models presented here represent the limit to what can be
achieved with single-component models of the outer atmospheres of late-type 
stars, and the single-dimensional 
radiative transfer calculations associated therewith; such models can account
for most of the observed properties of ultraviolet emission lines, but 
modern spectra are of sufficient quality to indicate some inadequacies. 
Evidence of inhomogeneity comes as no surprise; direct observation of the
Sun shows us that stellar atmospheres contain structure. However, at present
we have no direct means of observing inhomogeneity in other main-sequence
stars and so it can only be understood by modelling spectra.

Theoretical studies involving two-component models are already underway
(e.g. Cuntz et al. 1999). In the future constraints on modelling of this type
must be provided by semi-empirical multi-component modelling of spectra
involving multi-dimensional radiative transfer calculations (codes capable
of such multi-dimensional calculations are now available e.g. Uitenbroek 
2001).

The 
indications of inhomogeneity in the upper chromosphere/lower transition
region of $\epsilon$~Eri dovetail with our analysis of the upper transition
region/corona of the same star (Sim \& Jordan 2003b), in which the filling
factor of emitting material in the mid-transition region was found to be
$\sim 20$ per cent. Given the current absence of a definitive 
understanding of the heating mechanisms which operate in the lower transition
region and below, a similar argument cannot be applied to directly estimate
the filling factor in these parts of the atmosphere, but if the filling factor
is associated with magnetic fields (as expected) then solar observations 
would suggest that the filling factor should become smaller
deeper in the atmosphere. However, a full investigation of this must await 
quantitative 3-D modelling of the optically thick lines 
which can provide information about the
inhomogeneity; it is to be hoped that 
in the near future
such modelling 
can be combined 
with advances in theoretical studies
to lead us to  
a clearer understanding of the structure and heating of the
outer atmospheres of late-type stars.

\section*{Acknowledgements}

The majority of this work was carried out while SAS was a PPARC
supported graduate student at the University of Oxford 
(PPA/S/S/1999/02862). It was completed
while he was a PPARC supported 
PDRA at Imperial College London (PPA/G/S/2000/00032).

Some of the data presented in this paper were obtained from the 
Multimission Archive at the Space Telescope Science Institute (MAST). STScI 
is operated by the Association of Universities for Research in Astronomy, Inc.,
under NASA contract NAS5-26555. Support for MAST for non-HST data is provided 
by the NASA Office of Space Science via grant NAG5-7584 and by other grants 
and contracts.

\section*{Bibliography}

Abia C., Rebolo R., Beckman J. E., Crivellari L., 1988, \\
\indent A\&A, 206, 100\\
Arnaud M., Rothenflug R., 1985, A\&AS, 60, 425\\
Arnaud M., Raymond J., 1992, ApJ, 398, 394\\
Asplund M., Grevesse N., Sauval A. J., astro-ph/0410214\\
Bell K. L., Gilbody H. B., Hughes J. G., Kingston A. E.,\\
\indent Smith F. J., 1983, J. Phys. Chem. Ref. Data., 12, 891\\
Blum R. D., Pradhan A. K., 1992, ApJS, 80, 425\\
Bodaghee A., Santos N. C., Israelian G., Mayor M., 2003,\\
\indent A\&A, 404, 715\\
Burgess A., 1964, ApJ, 139, 776\\
Burgess A., 1965, in E.H. Avrett, O.J. Gingerich, \& C.A.\\
\indent Whitney eds., in proc. Second Harvard-Smithsonian\\
\indent Conference on Stellar Atmospheres, SAO Special\\ 
\indent Report No. 167, Smithsonian Institution\\
\indent Astrophysical Observatory, p.47\\
Burton W. M., Jordan C., Ridgeley A., Wilson R., 1971, \\
\indent Phil. Trans. Roy. Soc., London, A270, 81\\ 
Burton W. M., Jordan C., Ridgeley A., Wilson R., 1973, \\
\indent A\&A, 27, 101\\
Cardelli J. A., Meyer D. M., Jura M., Savage B. D., 1996,\\
\indent ApJ, 467, 334\\
Carlsson M., 1986, Uppsala Astronomical Observatory, \\
\indent Report No. 33\\
Crawford H. J., Price P. B., Sullivan J. D., 1972, ApJL,\\
\indent 175, L149\\
Cunto W., Mendoza C., Ochsenbein F., Zeippen C. J.,\\
\indent 1993, A\&A, 275, L5\\
Cuntz M., Rammacher W., Ulmschneider P., Musielak Z.\\
\indent E., Saar S. H., 1999, ApJ, 552, 1053\\
Dere K. P., Landi E., Young P. R., Del Zanna G., 2001,\\
\indent ApJS, 134, 331\\
Dere K. P., Landi E., Mason H. E., Monsignori Fossi B. C.,\\
\indent Young P. R., 1997, A\&AS, 125, 149\\
Drake J. J., Smith G., 1993, ApJ, 412, 797\\
Drake J. J., Laming J. M., Widing K. G., 1995, ApJ, 443,\\
\indent 393\\
Drake J. J., Laming J. M., Widing K. G., 1997, ApJ, 478,\\
\indent 403\\
Dring A. R., Linsky J., Murthy J., Henry R. C., Moos W., \\
\indent Vidal-Madjar A., Audouze J., Landsman W., 1997,\\
\indent ApJ, 488, 760\\
Dufton P. L., Kingston A. E., 1991, MNRAS, 248, 827\\
Dupree A. K., 1972, ApJ, 178, 527\\
Feldman U., Mandelbaum P., Seely J. F., Doschek G. A., \\
\indent Gursky H., 1992, ApJS, 81, 387 \\
Fontenla J. M., Avrett E. H., Loeser R., 1993, ApJ, 406,\\
\indent 319\\
Fontenla J. M., Avrett E. H., Loeser R., 2002, ApJ, 572,\\
\indent 636\\
Goldberb L., Dupree A. K., Allen J. W., 1965, Annales\\
\indent d'Astrophysique, 28, 589\\
Grevesse N., Sauval A. J., 1998, Space Sci. Rev., 85, 161\\
Griesmann U., Kling R., 2000, ApJL, 536, L113\\
Griffiths N. W., Jordan C., 1998, ApJ, 497, 883\\
Harper G. M., 1992, MNRAS, 256, 37\\
Hubeny I., Lites B. W., 1995, ApJ, 455, 376\\ 
Jordan C., 1969a, MNRAS, 142, 501\\
Jordan C., 1969b, ApJ, 156, 49\\
Jordan C., Brown A., 1981 in R. M. Bonnet \& A. K. Dupree\\
\indent eds., Solar Phenomena in Stars and Stellar Systems, \\
\indent NATO ASIC, 68, Reidel. Dordrecht, Holland, p.199\\
Jordan C., Macpherson K. P., Smith G. R., 2001, MNRAS \\
\indent 328, 1098 \\ 
Jordan C., Smith G. R., Houdebine E. R., 2005, MNRAS \\
\indent submitted \\
Jordan C., McMurry A. D., Sim S. A., Arulvel M., 2001b, \\
\indent MNRAS, 322, L5 \\
Jordan C., Sim S. A., McMurry A. D., Arulvel M., 2001a \\
\indent MNRAS, 326, 303 \\
Jordan C., Ayres T. R., Brown A., Linsky J. L., Simon T.,\\
\indent 1987, MNRAS, 225, 903\\
Jordan C., Doschek G. A., Drake, J. J., Galvin A. B.,\\
\indent  Raymond J. C., 1998, in R. A. Donahue \& \\
\indent J. A. Bookbinder eds., ASP Conf. Ser. 154: Cool\\
\indent Stars, Stellar Systems, and the Sun, 10, p. 91\\
Judge P. G., Brekke P., 1994, in eds. K. S. \\
\indent Balasubramaniam \& G. Simon, Proc. 14th \\
\indent International Summer Workshop: Solar Active Region \\
\indent Evolution -- Comparing Models with Observations, San \\
\indent Francisco: ASP, p. 321 \\
Judge P. G., Carlsson M., Stein R. F., 2003, ApJ, 597, 1158\\
Judge P. G., Hubeny V., Brown J. C., 1997, ApJ, 475, 275\\
Judge P. G., Woods T. N., Brekke P., Rottman G. J. 1995,\\
\indent ApJ, 455, L85\\
Kaufman V., Martin W. C., 1993, J. Phys. Chem. Ref. Data, 22\\
\indent 279\\
Kelch W. L., 1978, ApJ, 222, 931\\
Kurucz R. L., Bell B., 1995, Atomic Line Data CD-ROM\\
\indent No. 23, Cambridge, Mass: S. A. O.\\
Laming J. M., Drake J. J., 1999, ApJ, 516, 324\\
Laming J. M., Drake J. J., Widing K. G., 1995, ApJ, 443,\\
\indent 416\\
Laming J. M., Drake J. J., Widing K. G., 1996, ApJ, 462,\\
\indent 948\\
McIntosh S. W., Brown J. C., Judge P. G., 1998, A\&A,\\
\indent 333, 333\\
McMurry A. D., Jordan C., 2000, MNRAS, 313, 423\\
Mazzotta P., Mazzitelli G., Colafrancesco S., Vittorio N.,\\
\indent 1998, A\&AS, 133,403\\
Nahar S. N., 1995, ApJS, 101, 423\\
Nahar S. N., 1996, ApJS, 106, 213\\
Nahar S. N., 1999, ApJS, 120, 131\\
Nahar S. N. Pradhan A. K., 1997, ApJS, 111, 339\\
Nahar S. N., Pradhan A. K., Zhang H. L., 1995, ApJS, 131,\\
\indent 375\\
Pietarila A., Judge P. G., 2004, ApJ, 606, 1239\\
Redfield S., Linsky J. L., 2002, ApJS, 139, 439\\
Redfield S., Linsky J. L., Ake T. B., Ayres T. R., Dupree\\
\indent A. K., Robinson R. D., Wood B. E., Young P. R.,\\
\indent 2002, ApJ, 581, 626\\
Sankrit R., Kruk J. W., Ake T. B., Andersson B.-G., 2001, \\
\indent FUSE Data Analysis Cookbook, Version 1.0\\
Scharmer G. B., Carlsson M., 1985a, J. Comput. Phys., 59,\\
\indent 56\\
Scharmer G. B., Carlsson M., 1985b, in J. Beckman \& L.\\
\indent  Crivellari eds., Progress in Stellar Spectral Line\\
\indent Formation Theory, NATO ASI Series C, 152, p. 189\\
Shine R. A., 1983, ApJ, 266, 882\\
Simon T., Kelch W. L., Linsky J. L., 1980, ApJ, 237, 72\\
Sim S. A., 2001, MNRAS, 326, 821\\
Sim S. A., 2002, D.Phil Thesis, University of Oxford\\
Sim S. A., Jordan C., 2003a, MNRAS, 341, 517\\
Sim S. A., Jordan C., 2003b, MNRAS, 346, 846\\
Summers H. P., 1972, MNRAS, 158, 255\\
Summers H. P., 1974a, MNRAS, 169, 663\\
Summers H. P., 1974b, Culham Lab. Tech. Rep., 367\\
Thatcher J. D., Robinson R. D., Rees D. E., 1991, MNRAS,\\
\indent 250, 14\\
Uitenbroek H., 1989, A\&A, 213, 360\\
Uitenbroek H., 2001, ApJ, 557, 389\\
Vernazza J. E., Avrett E. H., Loeser R., 1973, ApJ, 184,\\
\indent 605\\
Vernazza J. E., Avrett E. H., Loeser R., 1976, ApJS, 30, 1\\
Vernazza J. E., Avrett E. H., Loeser R., 1981, ApJS, 45,\\
\indent 635\\
Wood B. E., Ambruster C. W., Brown A., Linsky J. L.,\\
 \indent ApJ, 542, 411\\
Zhao G., Chen Y. Q., Qiu H. M., Li Z. W., 2002, ApJ, 124,\\
\indent 2224\\

\newpage
\section*{Appendix A}

Tables A1 -- A4 give the computed logarithmic ionization fractions
for C, N O and Si, respectively. See Section~3 for details of these
calculation.

\section*{Appendix B}

Tables B1 -- B3 give parameters for the atmospheric models A, B and C 
 (as presented in Sections 5 and 6, respectively).

\vspace{12cm}

\begin{table}
{{\bf A1.} Logarithmic ionization fractions computed for C (see Section~3).}\\
\begin{tabular}{lccc} \hline
$\log T_{e}(\mbox{K})$ & \multicolumn{3}{c}{Ionization fraction} \\
& C~{\sc ii} & C~{\sc iii} & C~{\sc iv} \\ \hline
4.00& -1.45&  --&  --\\
4.10& -0.34&  --&  --\\
4.20& -0.04& -4.05&  --\\
4.30& -0.01& -2.41&  --\\
4.40& -0.03& -1.20&  --\\
4.50& -0.18& -0.48& -4.72\\
4.60& -0.52& -0.16& -3.00\\
4.70& -0.97& -0.06& -1.74\\
4.80& -1.46& -0.09& -0.83\\
4.90& -2.07& -0.32& -0.31\\
5.00& -2.97& -0.89& -0.24\\
5.10& -4.08& -1.75& -0.62\\
5.20& -5.29& -2.72& -1.21\\
5.30& -6.41& -3.60& -1.76\\
5.40& -7.33& -4.32& -2.22\\
5.50& --& -4.89& -2.56\\ \hline
\end{tabular}
\end{table}
 
\begin{table}
{{\bf A2} Logarithmic ionization fractions computed for N (see Section~3).}\\
\begin{tabular}{lccc} \hline
$\log T_{e}(\mbox{K})$ & \multicolumn{3}{c}{Ionization fraction} \\
& N~{\sc iii} & N~{\sc iv} & N~{\sc v} \\ \hline
4.60& -0.54& -3.88&  --\\
4.70& -0.18& -2.33&  --\\
4.80& -0.09& -1.29& -4.55\\
4.90& -0.15& -0.59& -2.69\\
5.00& -0.42& -0.25& -1.41\\
5.10& -0.89& -0.21& -0.60\\
5.20& -1.61& -0.52& -0.30\\
5.30& -2.65& -1.19& -0.50\\
5.40& -3.79& -2.03& -0.97\\
5.50& -4.85& -2.85& -1.48\\ \hline
\end{tabular}
\end{table}
 
\begin{table}
{{\bf A3} Logarithmic ionization fractions computed for O (see Section~3).}\\
\begin{tabular}{lcccc} \hline
$\log T_{e}(\mbox{K})$ & \multicolumn{4}{c}{Ionization fraction} \\
&O~{\sc iii} & O~{\sc iv} & O~{\sc v} & O~{\sc vi} \\ \hline
4.50& -1.83&  --&  --&  --\\
4.60& -0.77& -4.64&  --&  --\\
4.70& -0.24& -2.74&  --&  --\\
4.80& -0.09& -1.53& -5.27&  --\\
4.90& -0.11& -0.77& -3.37&  --\\
5.00& -0.30& -0.32& -1.99&  --\\
5.10& -0.66& -0.16& -1.06& -3.25\\
5.20& -1.16& -0.23& -0.49& -1.76\\
5.30& -1.80& -0.53& -0.28& -0.81\\
5.40& -2.66& -1.11& -0.44& -0.41\\
5.50& -3.84& -2.04& -1.03& -0.55\\ \hline
\end{tabular}
\end{table}

\begin{table}
{{\bf A4} Logarithmic ionization fractions computed for Si (see Section~3).}\\
\begin{tabular}{lccc} \hline
$\log T_{e}(\mbox{K})$ & \multicolumn{3}{c}{Ionization fraction} \\
& Si~{\sc ii} & Si~{\sc iii} & Si~{\sc iv} \\
4.00& -0.01& -1.74&  --\\
4.10& -0.03& -1.18&  --\\
4.20& -0.09& -0.71&  --\\
4.30& -0.33& -0.27& -4.23\\
4.40& -0.94& -0.05& -2.67\\
4.50& -1.60& -0.04& -1.27\\
4.60& -2.10& -0.18& -0.49\\
4.70& -2.53& -0.39& -0.24\\
4.80& -3.06& -0.69& -0.20\\
4.90& -3.92& -1.33& -0.53\\
5.00& -5.14& -2.36& -1.15\\
5.10& -6.28& -3.30& -1.67\\
5.20& -7.21& -4.02& -2.03\\
5.30& -8.16& -4.77& -2.48\\
5.40&  --& -4.86& -2.93\\
5.50&  --& -5.65& -3.45\\ \hline
\end{tabular}
\end{table}

\begin{table}
{{\bf B1} Model~A. $m_{col}$ is the column mass density, $T_{e}$ is the
electron temperature, $N_{e}$ is the electron number density,
$N_{H}$ is the hydrogen number density and $\xi_{\mbox{\scriptsize micro}}$ is the
most-probable microturbulent velocity.} \\
\begin{tabular}{lcccc} \hline
$\log m_{col}$ & $\log T_{e}$ & $\log N_{e}$ & $\log N_{H}$ & $\xi_{\mbox{\scriptsize micro}}$ \\ 
g cm$^{-2}$& K & cm$^{-3}$ & cm$^{-3}$ & km~s$^{-1}$\\ \hline
-4.50798&5.490&10.31&10.23&20.95\\
-4.50786&5.410&10.38&10.30&20.95\\
-4.50777&5.331&10.46&10.38&20.95\\
-4.50769&5.250&10.53&10.45&20.95\\
-4.50761&5.170&10.60&10.53&20.95\\
-4.50754&5.091&10.68&10.60&20.95\\
-4.50747&5.010&10.75&10.67&20.95\\
-4.50740&4.930&10.81&10.73&20.95\\
-4.50737&4.890&10.85&10.77&20.84\\
-4.50734&4.849&10.88&10.81&20.40\\
-4.50730&4.810&10.92&10.84&19.98\\
-4.50727&4.769&10.96&10.88&19.54\\
-4.50723&4.731&10.99&10.91&19.13\\
-4.50719&4.689&11.03&10.95&18.69\\
-4.50714&4.650&11.06&10.99&18.27\\
-4.50708&4.610&11.10&11.02&17.84\\
-4.50701&4.570&11.14&11.06&17.41\\
-4.50693&4.530&11.17&11.09&16.98\\
-4.50682&4.490&11.21&11.13&16.56\\
-4.50670&4.450&11.24&11.17&16.12\\
-4.50655&4.410&11.27&11.21&15.70\\
-4.50636&4.370&11.29&11.25&15.27\\
-4.50613&4.330&11.32&11.29&14.84\\
-4.50585&4.290&11.35&11.33&14.41\\
-4.50550&4.250&11.37&11.37&13.98\\
-4.50507&4.210&11.39&11.41&13.55\\
-4.50453&4.170&11.42&11.46&13.12\\
-4.50381&4.130&11.44&11.50&12.70\\
-4.50278&4.090&11.45&11.55&12.27\\
-4.50116&4.050&11.44&11.60&11.84\\
-4.49819&4.010&11.43&11.66&11.41\\
-4.49184&3.970&11.41&11.72&10.98\\
-4.47421&3.930&11.34&11.80&10.55\\
-4.40442&3.890&11.18&11.93&10.12\\
-4.15000&3.855&11.04&12.24& 9.75\\
-3.86000&3.835&11.02&12.56& 9.53\\
-3.49000&3.815&11.05&12.95& 9.32\\
-3.13000&3.785&11.02&13.39& 8.01\\
-2.80000&3.748&10.89&13.83& 6.27\\
-2.41000&3.712&10.81&14.32& 4.76\\
-2.06000&3.678&10.85&14.76& 3.41\\
-1.80000&3.651&11.03&15.08& 2.40\\
-1.69000&3.639&11.12&15.21& 1.97\\
-1.57000&3.626&11.21&15.35& 1.56\\
-1.43000&3.619&11.31&15.51& 1.31\\
-1.27000&3.617&11.43&15.67& 1.25\\
-1.06000&3.618&11.59&15.88& 1.25\\
-0.86000&3.619&11.74&16.08& 1.25\\
-0.61000&3.619&11.92&16.33& 1.25\\
-0.39000&3.619&12.08&16.54& 1.25\\
-0.15000&3.621&12.25&16.78& 1.25\\
 0.06000&3.626&12.43&16.99& 1.25\\
 0.26000&3.641&12.66&17.17& 1.25\\
 0.43000&3.666&12.93&17.32& 1.25\\
 0.59893&3.704&13.24&17.45& 1.25\\
 0.70717&3.736&13.44&17.53& 1.25\\
 0.79292&3.763&13.65&17.59& 1.25\\
 0.87069&3.788&13.91&17.64& 1.25\\ \hline
\end{tabular}
\end{table}

\begin{table}
{{\bf B2} Model~B. $m_{col}$ is the column mass density, $T_{e}$ is the
electron temperature, $N_{e}$ is the electron number density,
$N_{H}$ is the hydrogen number density and $\xi_{\mbox{\scriptsize micro}}$ is the
most-probable microturbulent velocity.} \\
\begin{tabular}{lcccc} \hline
$\log m_{col}$ & $\log T_{e}$ & $\log N_{e}$ & $\log N_{H}$ & $\xi_{\mbox{\scriptsize micro}}$ \\ 
g cm$^{-2}$& K & cm$^{-3}$ & cm$^{-3}$ & km~s$^{-1}$\\ \hline
-4.50517&5.490&10.31&10.25&19.15\\
-4.50505&5.410&10.39&10.32&19.15\\
-4.50494&5.330&10.46&10.40&19.15\\
-4.50483&5.250&10.54&10.47&19.15\\
-4.50472&5.169&10.61&10.55&19.15\\
-4.50463&5.089&10.69&10.62&19.15\\
-4.50455&5.009&10.76&10.69&19.15\\
-4.50450&4.949&10.81&10.75&19.15\\
-4.50446&4.910&10.84&10.78&19.15\\
-4.50443&4.871&10.88&10.82&18.80\\
-4.50439&4.829&10.92&10.85&18.31\\
-4.50436&4.790&10.96&10.89&17.85\\
-4.50432&4.750&10.99&10.93&17.37\\
-4.50429&4.710&11.03&10.97&16.90\\
-4.50425&4.671&11.07&11.00&16.42\\
-4.50420&4.630&11.11&11.04&15.93\\
-4.50414&4.590&11.14&11.08&15.44\\
-4.50406&4.550&11.18&11.12&14.95\\
-4.50397&4.510&11.22&11.15&14.46\\
-4.50386&4.470&11.26&11.19&13.96\\
-4.50373&4.430&11.29&11.24&13.45\\
-4.50358&4.390&11.32&11.28&12.94\\
-4.50345&4.350&11.35&11.33&12.42\\
-4.50330&4.310&11.38&11.37&11.90\\
-4.50311&4.270&11.42&11.41&11.37\\
-4.50282&4.230&11.45&11.46&10.83\\
-4.50238&4.190&11.48&11.51&10.28\\
-4.50164&4.150&11.51&11.56& 9.72\\
-4.50037&4.110&11.54&11.61& 9.14\\
-4.49767&4.070&11.57&11.67& 8.55\\
-4.49026&4.030&11.59&11.73& 7.93\\
-4.47013&3.990&11.60&11.83& 7.29\\
-4.43521&3.950&11.55&11.95& 6.61\\
-4.37097&3.910&11.45&12.11& 5.89\\
-4.20254&3.870&11.29&12.38& 5.10\\
-3.90000&3.838&11.19&12.75& 4.40\\
-3.60000&3.813&11.14&13.09& 3.78\\
-3.25000&3.785&11.08&13.48& 3.31\\
-2.99000&3.764&11.02&13.77& 3.01\\
-2.61000&3.731&10.92&14.20& 2.56\\
-2.22000&3.694&10.87&14.63& 2.10\\
-2.06000&3.678&10.90&14.81& 1.91\\
-1.92000&3.664&10.97&14.97& 1.74\\
-1.80000&3.651&11.06&15.11& 1.60\\
-1.69000&3.639&11.14&15.23& 1.47\\
-1.57000&3.626&11.22&15.37& 1.34\\
-1.43000&3.619&11.31&15.51& 1.27\\
-1.27000&3.617&11.43&15.68& 1.25\\
-1.06000&3.618&11.60&15.89& 1.25\\
-0.86000&3.619&11.75&16.08& 1.25\\
-0.61000&3.619&11.93&16.33& 1.25\\
-0.39000&3.619&12.08&16.55& 1.25\\
-0.15000&3.621&12.26&16.79& 1.25\\
 0.06000&3.626&12.43&16.99& 1.25\\
 0.26000&3.641&12.66&17.18& 1.25\\
 0.43000&3.666&12.93&17.33& 1.25\\
 0.59893&3.704&13.24&17.46& 1.25\\
 0.70717&3.736&13.45&17.54& 1.25\\
 0.79292&3.763&13.66&17.60& 1.25\\
 0.87069&3.788&13.92&17.65& 1.25\\ \hline
\end{tabular}
\end{table}

\begin{table}
{{\bf B3} Model~C. $m_{col}$ is the column mass density, $T_{e}$ is the
electron temperature, $N_{e}$ is the electron number density,
$N_{H}$ is the hydrogen number density and $\xi_{\mbox{\scriptsize micro}}$ is the
most-probable microturbulent velocity.} \\
\begin{tabular}{lcccc} \hline
$\log m_{col}$ & $\log T_{e}$ & $\log N_{e}$ & $\log N_{H}$ & $\xi_{\mbox{\scriptsize micro}}$ \\ 
g cm$^{-2}$& K & cm$^{-3}$ & cm$^{-3}$ & km~s$^{-1}$\\ \hline
-4.54760&5.490&10.27&10.20&19.15\\
-4.54746&5.410&10.34&10.28&19.15\\
-4.54732&5.330&10.42&10.36&19.15\\
-4.54719&5.250&10.50&10.43&19.15\\
-4.54706&5.171&10.57&10.50&19.15\\
-4.54695&5.090&10.64&10.58&19.15\\
-4.54685&5.011&10.71&10.65&19.15\\
-4.54676&4.930&10.79&10.72&19.15\\
-4.54668&4.850&10.86&10.79&18.55\\
-4.54660&4.770&10.93&10.87&17.61\\
-4.54652&4.689&11.01&10.94&16.64\\
-4.54648&4.649&11.05&10.98&16.16\\
-4.54644&4.611&11.08&11.02&15.70\\
-4.54640&4.570&11.12&11.05&15.20\\
-4.54636&4.530&11.16&11.09&14.71\\
-4.54632&4.489&11.20&11.13&14.20\\
-4.54627&4.450&11.23&11.17&13.71\\
-4.54622&4.411&11.26&11.21&13.20\\
-4.54614&4.370&11.29&11.26&12.69\\
-4.54604&4.330&11.32&11.30&12.17\\
-4.54591&4.290&11.35&11.35&11.64\\
-4.54573&4.250&11.38&11.40&11.10\\
-4.54543&4.210&11.41&11.45&10.56\\
-4.54492&4.170&11.43&11.50&10.00\\
-4.54408&4.130&11.46&11.55& 9.43\\
-4.54267&4.090&11.48&11.61& 8.85\\
-4.54026&4.050&11.48&11.68& 8.24\\
-4.53602&4.010&11.48&11.75& 7.61\\
-4.52840&3.970&11.45&11.83& 6.95\\
-4.51317&3.930&11.38&11.93& 6.26\\
-4.45968&3.890&11.24&12.07& 5.50\\
-4.38000&3.870&11.13&12.20& 5.09\\
-4.26000&3.854&11.05&12.35& 4.75\\
-4.07000&3.838&11.00&12.58& 4.39\\
-3.85907&3.824&10.99&12.81& 4.06\\
-3.54154&3.806&11.00&13.16& 3.63\\
-3.21000&3.782&10.98&13.53& 3.27\\
-2.90422&3.757&10.94&13.87& 2.90\\
-2.57000&3.727&10.91&14.24& 2.51\\
-2.45000&3.716&10.91&14.38& 2.37\\
-2.22000&3.694&10.98&14.63& 2.10\\
-2.06000&3.678&11.09&14.81& 1.91\\
-1.92000&3.664&11.20&14.97& 1.74\\
-1.80000&3.651&11.30&15.11& 1.60\\
-1.69000&3.639&11.37&15.23& 1.47\\
-1.57000&3.626&11.44&15.37& 1.34\\
-1.43000&3.619&11.52&15.51& 1.27\\
-1.27000&3.617&11.63&15.68& 1.25\\
-1.06000&3.618&11.78&15.89& 1.25\\
-0.86000&3.619&11.93&16.08& 1.25\\
-0.61000&3.619&12.10&16.33& 1.25\\
-0.39000&3.619&12.25&16.55& 1.25\\
-0.15000&3.621&12.42&16.79& 1.25\\
 0.06000&3.626&12.59&16.99& 1.25\\
 0.26000&3.641&12.82&17.18& 1.25\\
 0.43000&3.666&13.11&17.33& 1.25\\
 0.59893&3.704&13.44&17.46& 1.25\\
 0.70717&3.736&13.64&17.54& 1.25\\
 0.79292&3.763&13.82&17.60& 1.25\\
 0.87069&3.788&14.02&17.65& 1.25\\ \hline
\end{tabular}
\end{table}

\label{lastpage}
\end{document}